\documentclass{jaa}

\usepackage{graphicx}

\usepackage{multirow}
\usepackage[fleqn]{amsmath}
\usepackage{lscape}
\usepackage{colortbl}
\usepackage{txfonts,balance,textcase,float}

\newcommand{\aap}   {{ Astron. Astrophys.}}
\newcommand{\aaps}  {{ Astron. Astrophys. Suppl.}}

\newcommand{\aj}    {{ Astron. J.}} 
\newcommand{\apj}   {{ Astrophys. J.}}

\newcommand{\mnras} {{ Mon. Not. Roy. Astron. Soc.}}

\newcommand{\pasp}  {{ Pub. Astron. Soc. Pac.}}

\newcommand{\na}    {{ New Astron.}} 

\newcommand{\actaa} {{ Acta Astron.}}

\newcommand{\etal}{{\em et al. }}

\begin{document}\sloppy

\title{$UBV(RI)_{KC}$ Photometry of NGC~2323 and NGC~2539 Open Clusters}


\author{{\.I}nci Akkaya Oralhan\textsuperscript{1,*}, Ra\'ul Michel\textsuperscript{2}, William J. Schuster\textsuperscript{2}, Y\"uksel Karata\c{s}\textsuperscript{3},Yonca Karsl\i\textsuperscript{1} and Carlos Chavarr\'{\i}a\textsuperscript{2}}
\affilOne{\textsuperscript{1}Erciyes University, Astronomy and Space Science Observatory, TR-38039, Kayseri, Turkey \\}
\affilTwo{\textsuperscript{2}Observatorio Astronomico Nacional, Universidad Nacional Autonoma de M\'exico, Apartado Postal 106, C.P. 22800, Ensenada, B.C., M\'exico. \\}
\affilThree{\textsuperscript{3}Department of Astronomy and Space Sciences, Science Faculty, Istanbul University,  34119, Istanbul, Turkey.\\}


\twocolumn[{

\maketitle

\corres{iakkaya@erciyes.edu.tr}

\msinfo{?? ?? 2019}{?? ?? 2019}

\begin{abstract}
The open clusters NGC~2323 and NGC~2539 have been analysed using CCD $UBV(RI)_{KC}$ photometric data, observed at the San Pedro M\'artir Observatory.  Cluster memberships have been determined with the proper motion and parallax measures from the GaiaDR2 astrometric data release. Photometric metal and heavy element abundances have been obtained as $([M/H],\;Z) = (-0.10,~0.012)$ (NGC~2323) and $(-0.31,\;0.007)$ (NGC~2539), from the $\delta(U$--$B)$ technique in the two colour diagrams, which are used to select the appropriate PARSEC isochrones.

The estimated reddening of NGC~2323 is E(B-V)=0.23$\pm$0.04 using 11 early type stars. For NGC~2539, we find   E(B-V)= 0.02$\pm$0.06. For $(B$--$V)$ colour, distance moduli and distances for NGC~2323 and NGC~2539 are derived as ($V_{0}$--$M_{\rm V}$, $d$ (pc)) = (10.00$\pm$0.10, 1000$\pm$50) and  ($V_{0}$--$M_{\rm V}$, $d$ (pc)) = (10.00$\pm$0.04, 1000$\pm$20), respectively. The median GaiaDR2 distance d=1000$\pm$140 pc ($\varpi$=0.998$\pm$0.136 mas) for the likely members of NGC~2323 is in concordance with its four colour photometric distances 910--1000 pc. For NGC 2539, its GaiaDR2 distance d=1330$\pm$250 pc ($\varpi$=0.751$\pm$0.139 mas) is close to its four colour photometric distances, 1000 pc.
 
Fitting the PARSEC isochrones to the colour magnitude diagrams (CMDs) gives an old age, 890$\pm$110 Myr, for NGC~2539. Whereas NGC~2323 has an intermediate-age, 200$\pm$50 Myr. One Red Clump/Red Giant candidate (BD-12 2380) in the CMDs of NGC~2539 has been confirmed as a member in terms of the distances $d_{I}$ =950$\pm$50 pc and $d_{V}$=910$\pm$90 pc of $VI$ filters within the uncertainties, as compared to the distance, 1000$\pm$20 pc of NGC~2539. This giant's GaiaDR2 distance (d=1200$\pm$70 pc) is not close to these photometric distances.
\end{abstract}


}]


\doinum{12.3456/s78910-011-012-3}
\artcitid{\#\#\#\#}
\volnum{000}
\year{0000}
\pgrange{1--}
\setcounter{page}{1}
\lp{1}

\section{Introduction}
CCD $U\!BV\!(RI)_{KC}$ photometric data which include the $U$ filter for open clusters (OCs) are quite valuable for determining the interstellar reddening, $E(B$--$V)$ and the photometric metallicity and heavy-element abundance ([M/H], Z) on colour-colour diagram (CC). The two parameters are vital importance for deriving the distance modulus, $(V_{0}$--$M_{\rm V})$ and the distance, $d$ (kpc) and age, $A$ (Gyr) from colour-magnitude diagrams (CMDs).

In this study we have analysed the deep CCD $U\!BV\!(RI)_{KC}$ photometry of the OCs, NGC~2323 and  NGC~2539 from the Sierra San Pedro M\'artir National Astronomical Observatory (SPMO) open cluster survey (cf. Schuster \etal 2007; Tapia \etal 2010 (T10); Akkaya \etal 2010 (A10); Akkaya Oralhan \etal 2015 (A15).
Both OCs are uniformly and homogeneously analysed with regards to the instrumentation, observing techniques, and reduction and  calibration methods. The two OCs which are located in 3rd quadrand  have no spectroscopic metal abundances in the literature. Their basic parameters from the literature (Kharchenko \etal 2013) are listed in Table 1. NGC 2323 has been studied by Sharma \etal (2006) using CCD UBVI photometry  and  by Claria \etal (1998) from photoelectric UBV photometry.  For NGC 2539, Choo \etal (2006) used CCD UBVI photometry to search variable stars. Joshi and Sagar (1986) studied its photoelectric UBV photometry. Claria and Lapsset (1986) use CCD DDO and CMT$_{1}$T$_{2}$ photometries for this OC. Many published works of these OCs (Table 11) are of the brighter stars, which are based on old photographic and photoelectric photometries.

Additionally, we use the GaiaDR2 astrometric data (proper motion components and parallaxes) (Gaia Collaboration, Brown \etal 2018) for determining the probable members for NGC~2323  and NGC~2539.  The membership determinations of previous works have been based on the proper motions ($PPMXL$) of Roeser \etal (2010) in combination with 2MASS JHK$_{s}$ values of Skrutskie \etal (2006). Cantat-Gaudin \etal (2018) state that the proper motion uncertainties of UCAC4 fall in the range of 1-10 mas~yr$^{-1}$ (Roeser \etal 2010; Zacharias \etal 2013), whereas the ones of Tycho-Gaia Astrometric Solution (TGAS) (Gaia Collaboration, Brown \etal 2016) have the range of 0.5-2.6 mas~yr$^{-1}$.

Because of the above-mentioned scientific grounds, we combine the recent GaiaDR2 astrometric data with deep CCD $U\!BV\!(RI)_{KC}$ photometry of both OCs, to derive their reddenings, metal abundances, distance moduli and distances, and ages for four colour indices, (B-V), (V-I), (R-I) and $(G_{BP}$--$G_{RP})$. This kind of photometric data are also valuable for classifying early-type stars, blue stragglers, and RC/RG candidates in the CMDs, and thus possible candidates are proposed for spectroscopic observations.

This paper is organized as follows:  Section~2 describes the observation and reduction techniques.  
The technique for determining cluster membership is presented in Section~3. Section~4 describes general procedures concerning the derivation of astrophysical parameters, and in Section~5 the results for these astrophysical parameters are presented. Discussions and Conclusions are given in Section~6.

\renewcommand{\arraystretch}{1.3} 
\begin{table}[!t]  
	\begin{center}
		\caption{The central equatorial (J2000) and Galactic coordinates plus the average 
			$\mu_{\alpha}$ and $\mu_{\delta}$ values together with their standard deviations of
			Kharchenko \etal (2013). Our findings from the GaiaDR2 data are given in parentheses.}
		\label{tab:coordinates-PM}
		{\scriptsize 
			\setlength{\tabcolsep}{0.51cm}
			\begin{tabular}{lrr}
				\hline
				Cluster &   NGC 2323 &   NGC 2539 \\
				\hline
				$\alpha_{2000}$ $(h\,m\,s)$ & 07 02 43.1 & 08 10 40.8\\
				& (07 02 51.6) &  (08 10 42) \\
				$\delta_{2000}$ $(^{\circ}\,^{\prime}\,^{\prime\prime})$ & $-$08 22 12 & $-$12 50 24 \\
				& ($-$08 20 42) & ($-$12 50 31) \\
				$\ell$ $(^{\circ})$  &     221.66 &     233.73 \\
				&     &     \\
				b$(^{\circ})$        &      -1.33 &      11.11 \\
				&          &      \\
				$\mu_{\alpha}$($mas~yr^{-1}$) & $+$0.50$\pm$ 0.17  & $-$2.28$\pm$ 0.19  \\
				& ($-$0.816$\pm$ 0.248) & ($-$2.347$\pm$ 0.181) \\
				$\mu_{\delta}$ ($mas~yr^{-1}$) & $-$1.75$\pm$ 0.17 &$-$1.91$\pm$ 0.19 \\
				& ($-$0.653$\pm$ 0.210) & ($-$0.569$\pm$0.205) \\
				\hline
			\end{tabular}
		}
	\end{center}
	
\end{table}

\begin{figure}[!t]\label{imagetwo.jpg}
	\begin{center}	
		\includegraphics[width=0.9\columnwidth]{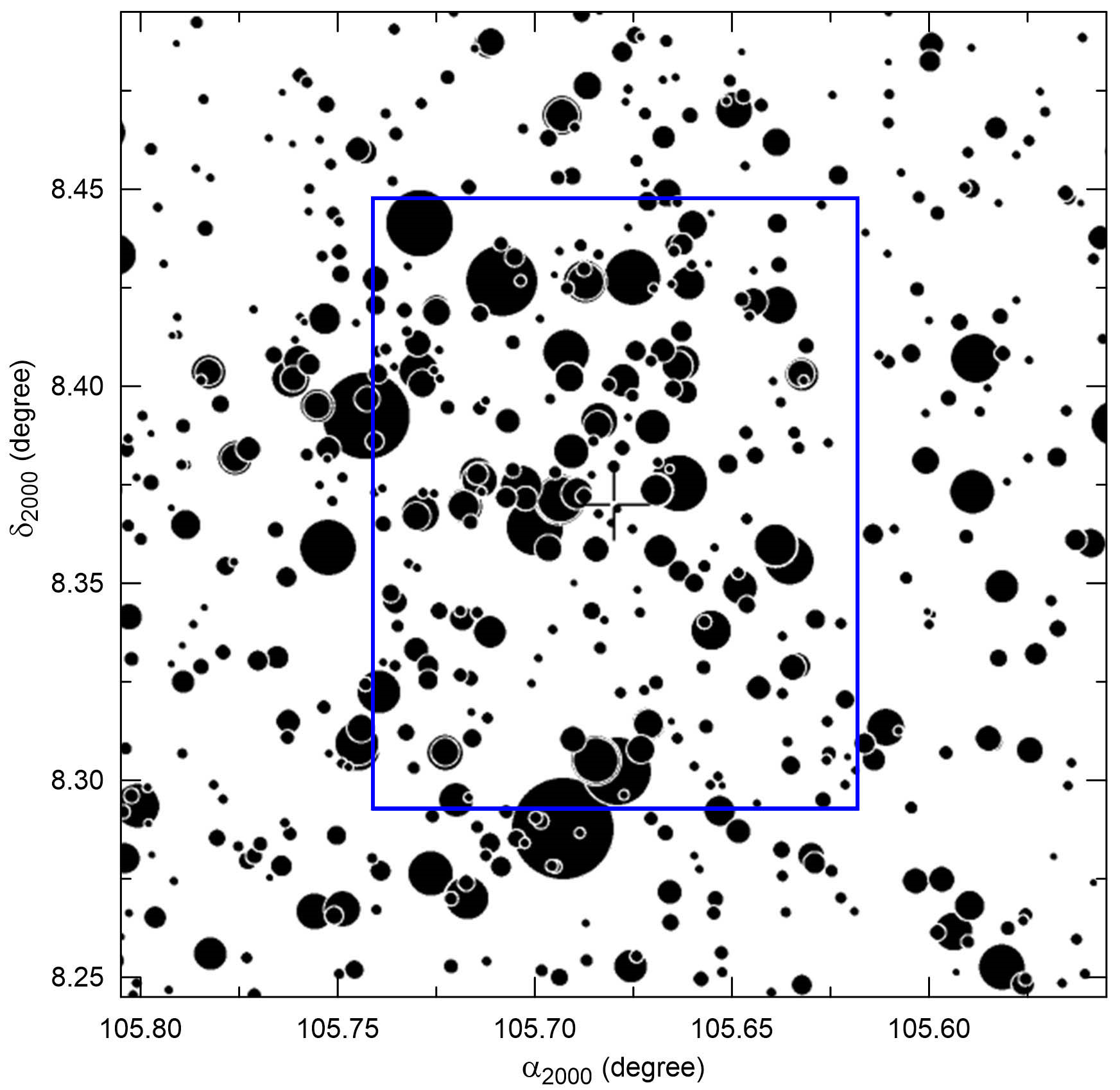}
		\includegraphics[width=0.9\columnwidth]{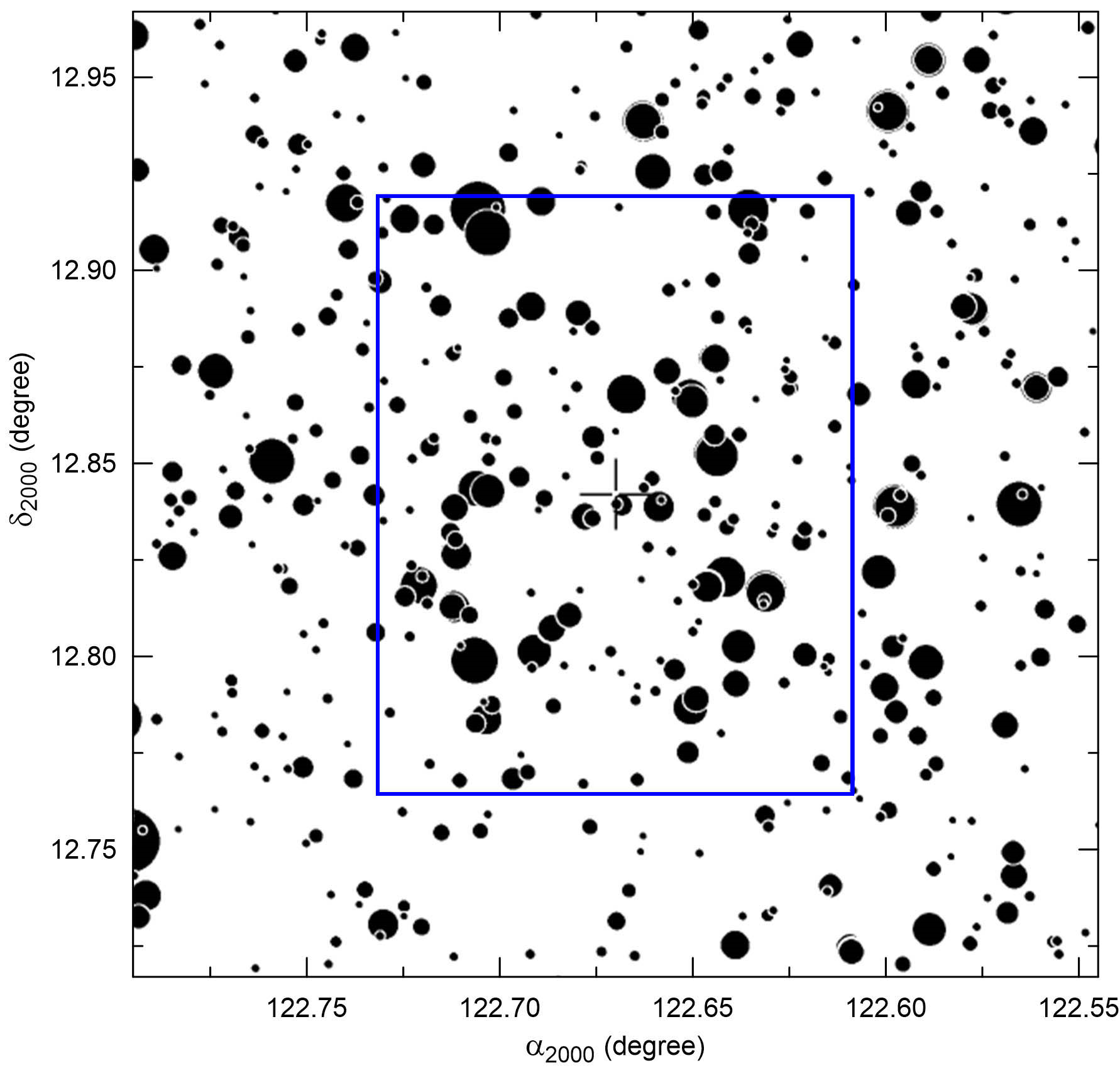}
		\caption{The images of https://www.aavso.org (AAVSO) of NGC~2323 ($18.3\,^\prime$x$17.23\,^\prime$)
			and NGC~2539 ($15.0\,^\prime$x $14.33\,^\prime$), are shown from top to bottom, respectively.  The
			blue rectangles indicate the field of view of the SPMO detector, $7.4^{\prime}\times9.3^{\prime}$.}
	\end{center}
\end{figure}

\begin{figure*}[!t]\label{pherr.jpg}
	\centerline{\includegraphics[width=1.5\columnwidth]{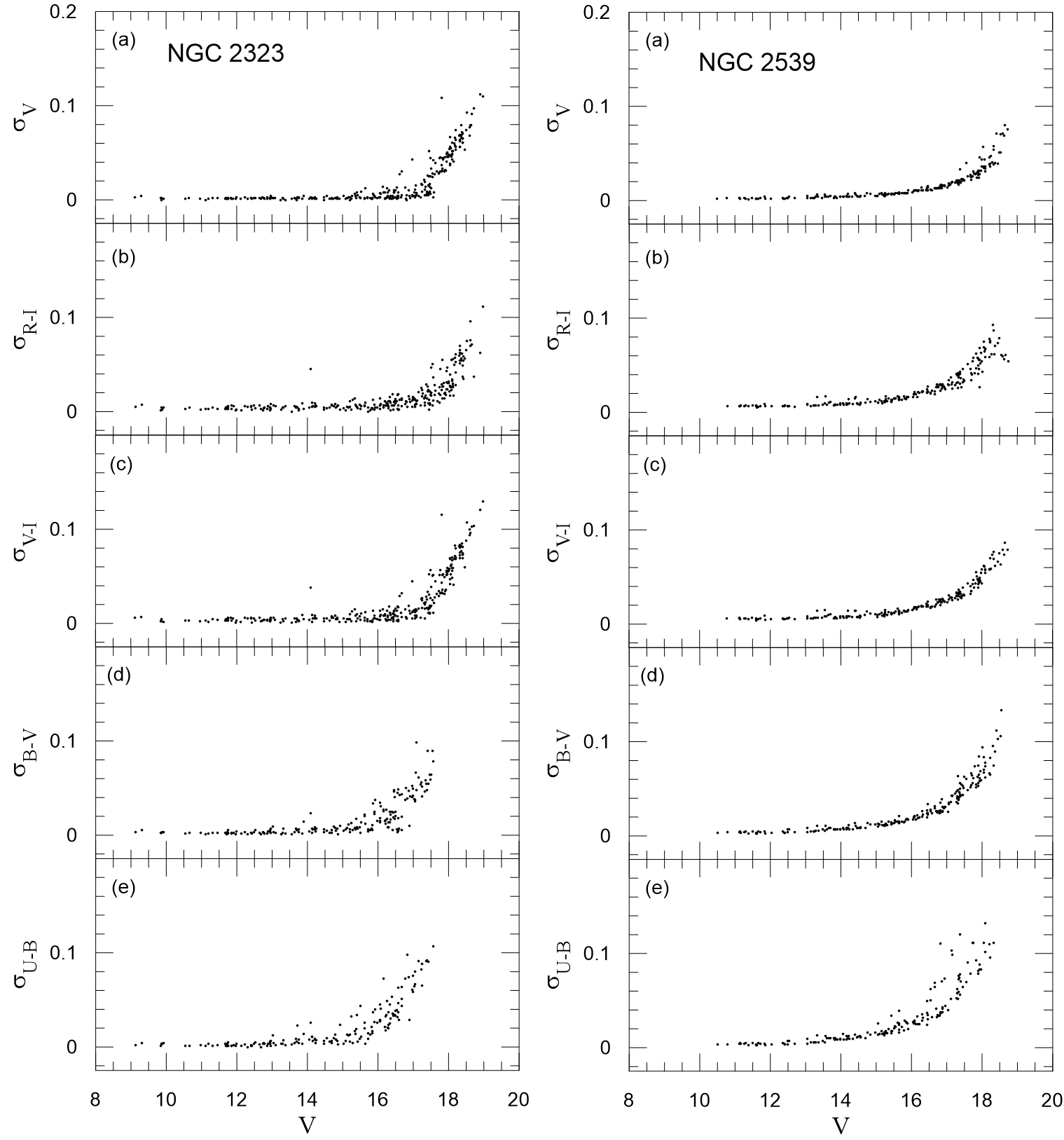}}
	\caption{Photometric errors for the $V$ magnitude and several broad-band colours plotted against
		the $V$ magnitudes of NGC~2323 and NGC~2539.}
\end{figure*}

\renewcommand{\arraystretch}{1.2} 
\begin{table*}[!t]  
	\caption{The observed open clusters and standard-star Landolt fields.}
	\label{tab:observation-log}
	{\scriptsize \tiny
		\setlength{\tabcolsep}{0.33cm}
		\begin{tabular}{lcccclllll}
			\hline
			Cluster& RA &DEC&LJD&Air mass&Filter $U$ & Filter $B$ & Filter $V$ & Filter $R$ & Filter $I$  \\
			&(2000) & (2000) &(days) &range& Exp. Time (s) & Exp. Time (s) & Exp. Time (s) & Exp. Time (s)      & Exp. Time (s)          \\
			\hline
			NGC 2323 & 07:02:52.4 & $-$08:21:01.2 & 2452669&1.352-1.382 & 60,300 & 10,30 & 6,20 & 4,15 & 4,15 \\
			NGC 2539 & 08:10:43.4 & $-$12:51:41.9 & 2452669&1.414-1.423 & 300    & 60    & 25& 10 &8          \\
			PG0918+029&09:21:32.6 & $+$02:47:42.9 & 2452669&1.135-1.136 & 300    & 220   &120& 80 & 80 \\
			RU 149    &07:24:15.5 & $-$00:32:58.0 & 2452669&1.174-1.175 &300     & 65    &45 & 50 & 50 \\
			RU 152    &07:29:58.4 & $-$02:06:11.2 & 2452669&1.198-1.523 &2x300   & 45,50 &25,27& 15,16 & 15,18 \\
			\hline
		\end{tabular}
	}
\end{table*}

\renewcommand{\arraystretch}{1.3} 
\begin{table*}[!t]  
	\caption{Coefficients of the transformation equations.}
	\label{tab:transf_coeffs_1}
	{\scriptsize \tiny
		\setlength{\tabcolsep}{0.37cm}
		\begin{tabular}{c|ccccc|ccccc}
			& \multicolumn{5}{c|}{F i r s t \qquad P a s s} & \multicolumn{5}{c}{S e c o n d \qquad P a s s} \\ 
			\hline
			index & $z$ & $k$ & $c$ & $rms$ & points & $c$ & $p$ & $q$ & $rms$ & points \\
			\hline
			UUB &  4.188$\pm$0.022 &  0.417$\pm$0.016 & -0.082$\pm$0.009 & 0.018 &	 50  & -0.144$\pm$0.050 &  0.031$\pm$0.039 &  0.010$\pm$0.015 & 0.022 &	 291  \\
			UUV &  4.198$\pm$0.024 &  0.428$\pm$0.017 & -0.049$\pm$0.006 & 0.020 &	 56  & -0.072$\pm$0.016 &  0.014$\pm$0.012 &  0.001$\pm$0.002 & 0.021 &	 301  \\
			BUB &  2.527$\pm$0.018 &  0.206$\pm$0.013 & -0.032$\pm$0.007 & 0.013 &	 50  & -0.104$\pm$0.055 &  0.014$\pm$0.043 &  0.047$\pm$0.015 & 0.023 &	 291  \\
			BBV &  2.571$\pm$0.019 &  0.194$\pm$0.013 & -0.052$\pm$0.010 & 0.014 &	 56  & -0.111$\pm$0.021 &  0.033$\pm$0.016 &  0.008$\pm$0.007 & 0.021 &	 355  \\
			VBV &  2.189$\pm$0.019 &  0.118$\pm$0.013 &  0.067$\pm$0.010 & 0.014 &	 56  &	0.033$\pm$0.020 &  0.012$\pm$0.015 &  0.012$\pm$0.007 & 0.020 &	 355  \\
			VVR &  2.170$\pm$0.021 &  0.131$\pm$0.015 &  0.122$\pm$0.022 & 0.017 &	 66  &	0.118$\pm$0.031 & -0.015$\pm$0.023 &  0.014$\pm$0.013 & 0.021 &	 427  \\
			RVR &  2.254$\pm$0.021 &  0.077$\pm$0.015 &  0.036$\pm$0.021 & 0.016 &	 66  & -0.002$\pm$0.032 &  0.013$\pm$0.025 &  0.029$\pm$0.014 & 0.022 &	 427  \\
			RRI &  2.226$\pm$0.014 &  0.091$\pm$0.011 &  0.065$\pm$0.010 & 0.010 &	 56  &	0.081$\pm$0.033 & -0.015$\pm$0.025 & -0.009$\pm$0.010 & 0.021 &	 388  \\
			IVI &  2.547$\pm$0.027 &  0.004$\pm$0.018 & -0.047$\pm$0.015 & 0.018 &	 52  & -0.084$\pm$0.019 &  0.036$\pm$0.014 & -0.010$\pm$0.003 & 0.024 &	 368  \\
			IRI &  2.532$\pm$0.028 &  0.021$\pm$0.020 & -0.116$\pm$0.017 & 0.022 &	 56  & -0.141$\pm$0.038 &  0.025$\pm$0.029 & -0.009$\pm$0.011 & 0.025 &	 388  \\
			\hline
		\end{tabular}
	}
\end{table*}

\renewcommand{\arraystretch}{1.4} 
\begin{table}[!t]  
	\caption{Observed standard stars (first three rows) 
		and the subset of extinction stars (three rows at the bottom).}
	{\tiny \tiny   
		\setlength{\tabcolsep}{0.44cm}
		\begin{tabular}{lll}
			\hline
			HJD   &Meas.$\&$(B-V)$\&$Air Mass & Landolt Fields \\
			\hline
			2452668 & 596;$-$0.284--1.605; 1.19--1.34 & PG1047+003 0035 CL13 \\ 
			&   &      RU149 0538 CL13 \\ 
			2452669 & 606;$-$0.284--1.727; 1.14--1.50 & G3061911464112566400RM \\
			&   &     PG0918+029 0068 CL13  \\ 
			&                       &PG1047+003 0035 CL13 \\
			&   &     RU149 0525 CL13 \\ 
			&   &      RU152 0561 CL13 \\ 
			\hline
			All     &1202;$-$0.284--1.727; 1.14--1.50 &  \\ 
			\hline
			
			&   &      \\ 
			\hline
			HJD   &Meas.$\&$(B-V)$\&$ Air Mass    & Landolt Fields \\
			\hline
			2452668 & 184;$-$0.034--1.170;1.19--1.34 & RU149 0557 CL13 \\ 
			2452669 &  96; 0.043--1.528;1.20--1.50 & G3061911464112566400RM \\ 
			&   &     RU152 0586 CL13  \\ 
			\hline		
			All     &  96;0.043--1.528; 1.20--1.50 &  \\ 
			\hline			
		\end{tabular}
	}
\end{table}

\section{Observations and reductions}
Observations were carried out at the SPMO, during photometric nights, on June 7-10, 2013 UT using the 0.84-m (f/15) Ritchey-Chretien telescope e\-quipped with the Mexman filter wheel and the ESOPO CCD detector, a 2048x4608 13.5-$\mu$m square-pixel e2v CCD42-90 with a gain of 1.7 e$^-$/ADU and a readout noise of 3.8 e$^-$ at 2$\times$2 binning. The combination of the telescope and detector ensures an unvignetted field of view of 7.4$\times$9.3 arcmin$^2$. Each OC was observed through Johnson's $UBV$ and Kron-Cousin's $RI$ \footnote {personel communication of A. Landolt.} filters with short and long exposure times in order to properly record both the brighter and fainter stars of the fields under study. Standard fields (Landolt 2009) were also observed near the meridian and at 1.14--1.52 air masses to properly determine the atmospheric extinction coefficients and the equations to transform the instrumental data to the standard system. Flat fields were taken nightly at the beginning and/or the end of the night, whereas biases were recorded between cluster observations. For more details see the works of T10 and  A10. The log of the observations is shown in Table~2. It contains the object names, coordinates at the centres of the observed fields, local Julian date of the observations, air-mass range during the observations, and exposure times in each band. The second part of Table~2 contains data of the observed standard fields.

The data reduction was carried out by R. Michel \footnote {Data may be requested from R. Michel.} using the IRAF\footnote {IRAF is distributed by the National Optical Observatories, operated by the Association of Universities for Research in Astronomy,	Inc., under  cooperative agreement with the National Science Foundation.} package and together with some home-made auxiliary Fortran programs and Awk scripts. The supervised-automatic procedure, implemented by C. Chavarria, can be condensed as follows. All the images were bias subtracted and flat-field corrected (CCDRED); cosmic rays were then removed with the L.A. Cosmic\footnote{http://www.astro.yale.edu/dokkum/lacosmic} script (van Dokkum 2001). 

The images of each set of observations were aligned and trimmed (IMALIGN), generating a template image of each cluster field (IMSUM). By using these template images, in addition to reference images from the ESO Digital Sky Survey\footnote {http://archive.eso.org/dss/dss}, and equatorial coordinates from the 2MASS All-Sky Catalog of Point Sources (Cutri \etal 2003), an astrometric solution is obtained (CCMAP and CCTRAN) providing a means to transform pixel coordinates to equatorial coordinates of each detected star.

For each image, the average sky level and its standard deviation are iteratively calculated with the help of the IMSTAT routine. The average $FWHM$ of the stars in each image is also found by means of DAOFIND, FITPSF, and filtering  routines. With this information, bright-unsaturated-isolated stars are identified and used iteratively to build the point spread function (PSF) for each image. Consequently, instrumental magnitudes are calculated using the ALLSTAR routine (Stetson 1987).

The aperture photometry (PHOT) of the standard stars was calculated using a fixed aperture radius of $2(FWHM+3\sigma)$ where $FWHM$ was the average $FWHM$ of the run, and $\sigma$ its standard deviation. The pixel coordinates of each measured star are then transformed to their corresponding equatorial coordinates. The standard magnitudes were taken from the catalogue by Landolt (2009) and supplemented with the secondary photometric standards by Cutri \etal (2013). As a result, the transformation coefficients were found (FITPARAMS) using Equations~1$-$10:
{\tiny
\begin{flalign}
	u &=  U + z_{UUB} + k_{UUB} X_{U} + c_{UUB} (U-B) + p_{UUB} X_{U} (U-B)\\[0pt]
	u &= U + z_{UUV} + k_{UUV} X_{U} + c_{UUV} (U-V) + p_{UUV} X_{U} (U-V)\\[0pt]
	b &= B + z_{BUB} + k_{BUB} X_{B} + c_{BUB} (U-B) + p_{BUB} X_{B} (U-B)\\[0pt]
	b &= B + z_{BBV} + k_{BBV} X_{B} + c_{BBV} (B-V) + p_{BBV} X_{B} (B-V)\\[0pt]
	v &= V + z_{VBV} + k_{VBV} X_{V} + c_{VBV} (B-V) + p_{VBV} X_{V} (B-V) \\[0pt]
	v &= V + z_{VVR} + k_{VVR} X_{V} + c_{VVR} (V-R) + p_{VVR} X_{V} (V-R)\\[0pt]
	r &= R + z_{RVR} + k_{RVR} X_{R} + c_{RVR} (V-R) + p_{RVR} X_{R} (V-R)\\[0pt]
	r &= R + z_{RRI} + k_{RRI} X_{R} + c_{RRI} (R-I) + p_{RRI} X_{R} (R-I) \\[0pt]
	i &= I + z_{IVI} + k_{IVI} X_{I} + c_{IVI} (V-I) + p_{IVI} X_{I} (V-I) \\[0pt]
	i &= I + z_{IRI} + k_{IRI} X_{I} + c_{IRI} (R-I) + p_{IRI} X_{I} (R-I),
\end{flalign}
}

\renewcommand{\arraystretch}{1.3} 
\begin{table}[!t]  
	\caption {The mean photometric errors of $V$, $(R$--$I)$, $(V$--$I)$, $(B$--$V)$, and $(U$--$B)$ 
		for NGC~2323 and  NGC~2539 in terms of the $V$ mag.}	
	\label{tab:photometric errors}
	\tiny
	\setlength{\tabcolsep}{0.42cm}
	\begin{tabular}{cccccc}
		\hline
		V &$\sigma_{V}$&$\sigma_{R-I}$&$\sigma_{V-I}$ &$\sigma_{B-V}$ &$\sigma_{U-B}$ \\
		\hline
		NGC~2323 &      &     &      &      &     \\
		\hline
		9-10 &0.002 &0.004 &0.004 &0.003 &0.003 \\
		10-11 &0.001 &0.003 &0.003 &0.002 &0.001 \\
		11-12 &0.001 &0.004 &0.004 &0.002 &0.002 \\
		12-13 &0.002 &0.004 &0.004 &0.003 &0.003 \\
		13-14 &0.002 &0.004 &0.004 &0.005 &0.007 \\
		14-15 &0.002 &0.006 &0.005 &0.005 &0.009 \\
		15-16 &0.003 &0.006 &0.006 &0.012 &0.018 \\
		16-17 &0.007 &0.009 &0.011 &0.025 &0.041 \\
		17-18 &0.024 &0.020 &0.030 &    - &   - \\
		18-19 &0.064 &0.050 &0.074 &    - &    - \\
		\hline
		&      &      &      &      &       \\
		NGC~2539 &   &  &       &         &        \\
		\hline
		10-11 &0.003 &0.006 &0.005 &0.004 &0.004 \\
		11-12 &0.002 &0.007 &0.006 &0.004 &0.004 \\
		12-13 &0.003 &0.006 &0.006 &0.005 &0.005 \\
		13-14 &0.004 &0.009 &0.008 &0.007 &0.008 \\
		14-15 &0.005 &0.010 &0.009 &0.009 &0.012 \\
		15-16 &0.007 &0.014 &0.012 &0.014 &0.021 \\
		16-17 &0.013 &0.024 &0.021 &0.025 &0.039 \\
		17-18 &0.022 &0.041 &0.037 &0.051 &0.077 \\
		18-19 &0.040 &0.067 &0.065 &0.085 &0.110 \\
		\hline	
	\end{tabular}  
\end{table}

\begin{figure*}[!t]\label{phcomp.jpg}
	\includegraphics[width=0.93\textwidth]{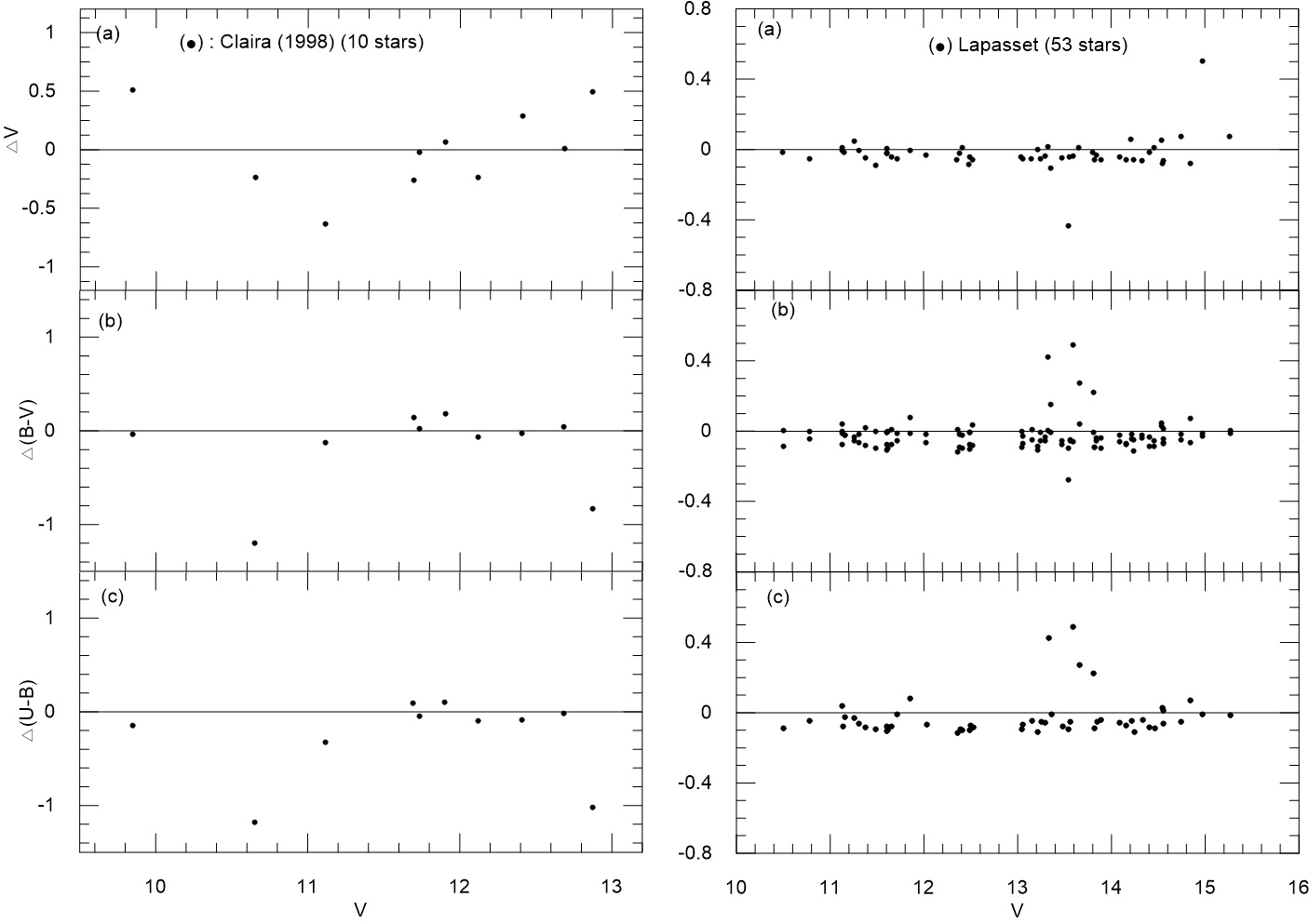}
	\caption{For NGC~2323 (left panels) and NGC~2539 (right panels), comparisons of the
		present CCD photometry with data from the literature:  Claria \etal (1998) and
		Lapasset \etal (2000), respectively.  The difference $\Delta$ (present photometry
		minus literature values) as a function of the $V$ mag.}
\end{figure*}

\begin{figure}[!t]\label{matchngc2323.jpg}
	\includegraphics[width=0.99\columnwidth]{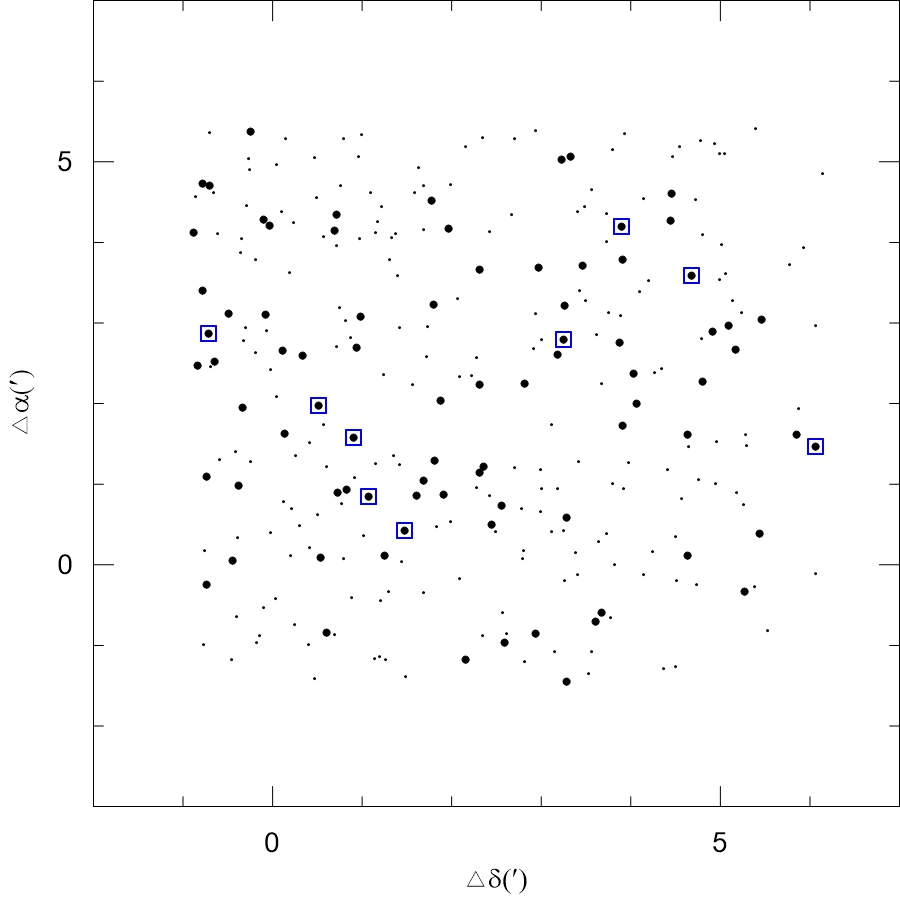}
	\caption{The finding chart is presented for NGC~2323 only.  The size of filled dots for our
		CCD~UBV photometry is proportional to the magnitudes of stars:  $V=9-15$ mag, large dots,
		and 15$-$19 mag, small dots. The blue squares show stars with photoelectric photometry
		of Claria \etal (1998) for $V < 13.00$.  $\Delta \alpha$ and $\Delta \delta$ are in units of
		arcmin, and are measured from the cluster center of NGC~2323 in WEBDA.}
\end{figure}

\noindent{where $u$, $b$, $v$, $r$, $i$ are the magnitudes in the instrumental system, $U$, $B$, $V$, $R$, $I$ the  magnitudes in the standard system, and $X$ is the air mass during the measurement. During the first iteration, only the first-order extinction measurements (for the stars observed over an extended range of air mass) were used to calculate the zero point ($z$), first-order extinction ($k$), and colour ($c$) coefficients; the second-order extinction ($p$) was zeroed. For the second interation, all data were used, having the zero points and first-order extinction values fixed to the values found in the previous iteration, while the colour-dependent coefficients ($p$ and $c$) were re-calculated. $p$ corrects the shift in $\lambda_{eff}$ due to the convolution of the spectral-energy distribution of the stars with the wavelength dependence of $k$. These values are given in Table 3.
	
For the case of this observing run, the number of valid measurements used in the transformations is shown in first three rows of Table~4 together with the colour and air-mass range near zenith of the standard fields. Table~4 includes the local Julian date of the observations, the number of measurements, the interval of the colour $(B$--$V)$, the air-mass range, and Landolt's field name. The resulting coefficients found using Equations~1$-$10, the $rms$ deviations of the fits, and the number of stars used in the calculations are given in three rows at the bottom of Table~4. New images that contained only the PSF stars were generated with the help of the GROUP, NSTAR and SUBSTAR routines, and aperture corrections were calculated with PHOT and MKAPFILE of IRAF. The mean value of the aperture correction is about $-0.2$ magnitude. 

Fig.~1 presents finding charts \footnote{Obtained from \textit{https://www.aavso.org (AAVSO)} web page of NGC~2323 ($18.3\,^\prime$x $17.23\,^\prime$) and NGC~2539 ($15.0\,^\prime$x $14.33\,^\prime$).} The blue rectangles indicate the analysed regions of these OCs, the field of view of the SPMO detector, $7.4^{\prime}\times9.3^{\prime}$.

The photometric errors in $V$ and colours $(R$--$I)$, $(V$--$I)$, $(B$--$V)$, $(U$--$B)$ of NGC~2323 (left panel) and NGC~2539 (right panel) are presented in Fig.~2. Their mean errors are listed in Table 5. Our inspection of Fig.~2 and Table~5 indicates that stars brighter than $V\approx17^m.00$ have errors smaller than $\approx0^m.05$ in both magnitudes and colours. For $V> 17^m.00$, large errors in $(U$--$B)$ dominate. The stars with $V<17^m.00$ have been considered for our analyses. 

A comparison of the present data with those in the literature for stars in common is done in Fig.~3 for NGC~2323 (left panels) and NGC~2539 (right panels). For the comparisons, we have considered only photoelectric photometry from the literature. For this purpose, a finding chart is shown in Fig.~4 for NGC~2323. The size of filled dots for our CCD~UBV photometry is proportional to the magnitudes of stars: $V=9^m.0$--$15^m.0$ (large dots) and $15^m.0$--$19^m.0$ (small dots). The blue squares show stars with the photoelectric photometry of Claria \etal (1998) for $V < 13^m.0$. From the left panels of Fig.~3 (for 10 stars in common), our magnitudes and colours, $V$, $(U$--$B)$, and $(B$--$V)$, do not show any systematic trends with the photolectric data of Claria \etal (1998). Three stars with up to $\Delta V$ = $0^m.50$ are seen in panel~(a), and two stars with large color deviations in $\Delta(U-B)$ and $\Delta(B-V)$, at $V=10^m.65$ and $V=12^m.87$, in panels (b)-(c) of Fig.~3. The magnitudes and colours of Claria \etal (1998) are somewhat fainter and redder than our CCD~UBV photometry.

For NGC~2539, the comparison with the photoelectric data of Lapasset (2000) is done for 53 stars in common in the right panels of Fig.~3. For $V=(13^m.0, 14^m.0)$, differences up to $0^m.40$ in $(B$--$V)$ and $(U$--$B)$ for a few stars indicate that our photometry gives bluer colours than the ones of Lapasset (2000); these 4--5 deviating stars are not easily explained, not by cosmic rays or cosmetic effects in the CCD data, but perhaps by stellar variability or typographical errors of Lapasset (2000).

\begin{figure*}[!t]\label{ngc2323pplcmd.jpg}
\begin {center}
\includegraphics[width=0.75\columnwidth]{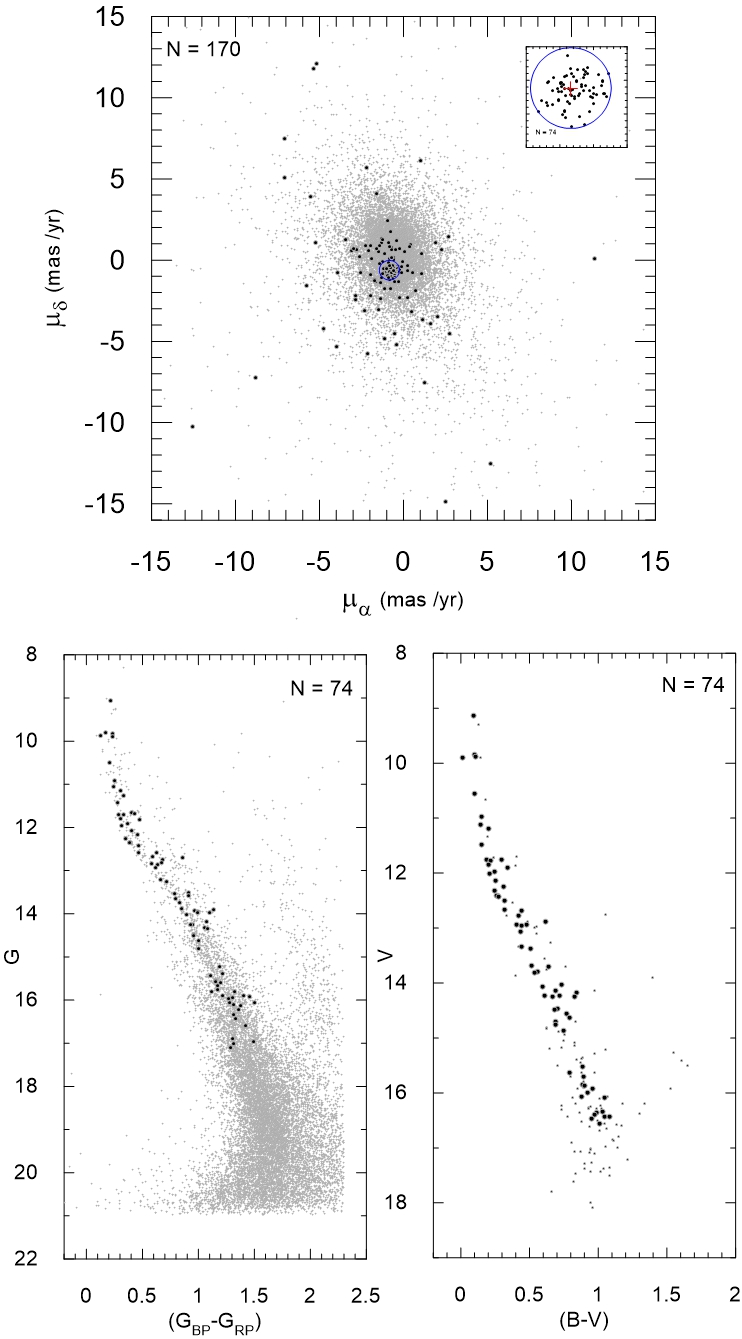}
\includegraphics[width=0.81\columnwidth]{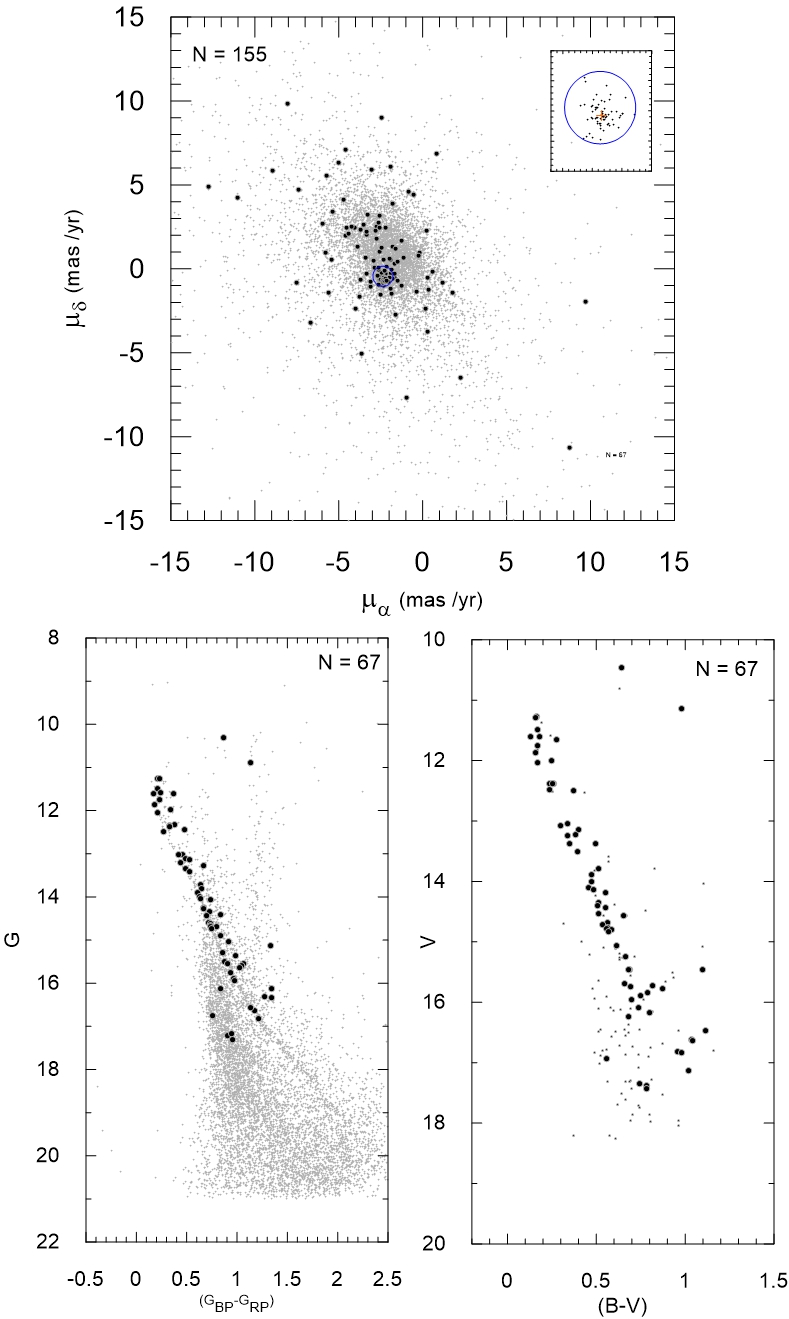}
\caption{The $\mu_{\alpha}$ versus $\mu_{\delta}$ for  NGC~2323 (left, top panel)
and NGC~2539 (right, top panel).  Small dots represent the GaiaDR2 astrometric and
photometric data for a 20 arcmin region centered on the two OCs.  Blue circles denote
the radii of 0.6 mas~yr$^{-1}$.  The likely members inside the blue circles are also
indicated in the inset of the top panels.  The big red pluses indicate the median
proper motions of the OCs.  The bottom panels show the $G$, $G_{BP}$--$G_{RP}$ of GaiaDR2 stars 
for a 20 arcmin field region which is centered on the OCs and the likely
cluster members (filled dots).  $V$, $(B$--$V)$ plots are also displayed for all
cluster stars:  170 small dots, NGC 2323, and 155 small dots, NGC 2539, together
with their likely members (filled dots) inside the blue circles. 
	}
	\end {center}
\end{figure*}

\section{Cluster Membership}
For the separation of the likely cluster members of NGC~2323 and NGC~2539, firstly, CCD $U\!BV\!(RI)_{KC}$ photometric data of NGC~2323 (170 stars) and NGC~2539 (155 stars) have been matched with the GaiaDR2 astrometric data (proper motions and parallaxes) from SIMBAD-VizieR\footnote{http://vizier.u-strasbg.fr/viz-bin/VizieR}. GaiaDR2 astrometric and photometric data of the field stars in a region of $R=20$ arcmin centered on our target OCs have also been considered to determine the cluster's radii. The $\mu_{\alpha}$ versus $\mu_{\delta}$ (Vector Point Diagram, VPD) of the two OCs are plotted in top panels of Fig.~5 for both the background/foreground field stars (small grey dots) and our OCs (filled dots). The proper motion radii (shown with blue circles) of 0.6 mas~yr$^{-1}$ around the centres of their VPDs define the membership criteria. These proper motion radii have been empirically fitted until the likely members inside these radii provide good single stellar sequence on $G$, $G_{BP}$--$G_{RP}$ and $V$, $(B$--$V)$ plots (bottom panels of Fig.~5). Therefore, the chosen radii are a good compromise for the likely cluster members. These proper motion radii have been constructed via the mathematical equations, $x = x_{0} + r\cos(\theta)$ and $y = y_{0} + r\sin(\theta)$. Here, (x$_{0}$, y$_{0}$) are the median values of ($\mu_{\alpha}$, $\mu_{\delta}$) (mas~yr$^{-1}$), the radius $r = \sqrt{\sigma_{\mu\alpha}^{2} + \sigma_{\mu\delta}^{2}}$ mas~yr$^{-1}$, and $\theta$ =$0^{\circ}$ to $360^{\circ}$.  Note that there almost seems to be a clear concentration of stars inside the chosen proper motion radii.  The inset plots (Fig.~5) show the likely cluster members inside the proper motion radii of NGC 2323 (74 probable members) and NGC 2539 (67 probable members). The big red pluses indicate the median values of proper motion components of the two OCs.

For the likely members of our OCs, the uncertainties of proper motions and the parallaxes are less than 0.30 mas~yr$^{-1}$ and 0.15 mas, respectively. These limits nearly remain within the uncertainties of the proper motion components of  $\sigma_{\mu\alpha}< 0.28$ mas~yr$^{-1}$ and $\sigma_{\mu\delta}< 0.24$  mas~yr$^{-1}$, and the uncertainty $\sigma_\varpi <0.16$ mas for G $<$ 18 mag,  which are reported by the Gaia Collaboration, Lindegren \etal (2018) (see their table B.1.).

The median proper motion components plus their equatorial coordinates, found from GaiaDR2 astrometric data for the likely members of our OCs are listed in the parentheses of Table 1. Having applied these proper motion criteria to our OCs, the numbers of the likely cluster members are now appropriate for determining the astrophysical parameters in CC and CMDs.

\section{Astrophysical parameters}

\subsection{Reddenings and photometric metal abundances}
The two colour $(U$--$B)$, $(B$--$V)$ diagrams of the probable members of NGC~2323 and NGC~2539 are displayed in Figs.~6 and 9.  Note that there are 56 (NGC 2323) and 42 stars (NGC 2539) which have $(U$--$B)$ colour.
It appears that NGC 2323 contains early-type stars, which are mostly members of young open clusters.  By using 11 early type stars with $(U-B)$ $<$ 0 (Fig.~6),  the mean interstellar reddening is estimated as $E(B-V)$ = 0.23$\pm$0.04 from Q--technique of early-type stars. The reddened intrinsic-colour sequence of the Schmidt-Kaler (SK82) (blue curve) for this $E(B-V)=0.23$ is fitted on the CC of NGC 2323.
For this, we adopt the following relations of Johnson and Morgan (1953) :  $Q=(U$--$B)-0.72(B$--$V)$, and $(B$--$V)_{0}=0.332Q$. Here E$(U-B) = 0.72E(B-V)+0.05E(B-V)^2$, $E(B$--$V)=(B$--$V)-(B$--$V)_{0}$ and $E(U$--$B)=(U$--$B)-(U$--$B)_{0}$. The interstellar reddening, $E(B-V)$ of NGC~2539 is derived from displacement of the intrinsic-colour sequences of dwarfs and red giants from Schmidt-Kaler (1982)(SK82) in the CC diagram (Fig.~9), until the best fit to the cluster members has been achieved: along the (U-B) axis by 0.72 E(B--V)+0.05E(B--V)$^2$ and along the $(B-V)$ axis by $E(B-V)$. 

F-type stars on the CC plots of these two OCs show ultraviolet excess, $\delta(U$--$B)$ above Hyades main sequence (MS) (green dashed curve), which are quite valuable for determining photometric metal abundance, $[M/H]$. Here, $\delta(U$--$B) = (U$--$B)_{Hyades}-\langle(U$--$B)_0\rangle$. By shifting the Hyades main-sequence  according to the E(B-V)'s (col.~2 of Table 6), a fit is applied to the F-type stars, the same as used to fit the RC/RG stars to the SK82 giant colours. The best fit of iso-metallicity curves as representative of the mean metal abundances of the two OCs are shown as red solid lines in CC diagrams (Figs.~6 and 9).
The average $\langle(B$--$V)_{0}\rangle$, $\langle(U$--$B)_{Hyades}\rangle$, $\langle(U$--$B)_{0}\rangle$ colours (cols.~3--5 of Table 6) have been fixed as mean values from the distribution of the F-type stars in each cluster. By using these average values, the ultraviolet excesses, $\delta(U$--$B)$ have been measured, and normalized to $(B$--$V)_0=0.6$ via the data of Table~1A given by Sandage (1969).
We use $[M/H]=+0.13(\pm0.04)-4.84(\pm0.60)\delta_{0.6}-7.93(\pm2.24)\delta_{0.6}^2$, by Karata\c{s} \& Schuster (2006) to estimate photometric metallicity values ([M/H]) of the two OCs. With the equation $Z = Z_\odot \cdot 10^{[M/H]}$, their [M/H] values are converted into the heavy-element abundance mass fraction, $Z$. The solar metal content is adopted as $Z_{\odot}=0.0152$. The mean values, $\delta(U$--$B)$, $\delta(U$--$B)_{0.6}$, $[M/H]$ and $Z$ of the two OCs are listed in cols.~6--9 of Table~6.

The $E(B$--$V)$ values from extinction maps given by Schlegel \etal (1998) (SFD) (based on the IRAS 100-micrometer surface brightness converted to extinction) have been obtained from NASA EXTRAGALACTIC DATABASE (NED) as $E(B$--$V)_{\rm SFD,\infty}=0.713$  and 0.032 for NGC~2323 and NGC~2539, respectively. Taking into consideration their distances d~(kpc) (Cols.~3 of Tables 7--8) and Galactic latitudes (Col.~5 of Table 1), the final reddening, $E(B$--$V)_{\rm SFD}$, for a given star is reduced compared to the total reddening $E(B$--$V)(\ell, b)_\infty$ by a factor $\lbrace1-\exp[-d \sin |b|/H]\rbrace$, given by Bahcall \& Soneira (1980), where $b$, $d$, and $H$ are the Galactic latitude, distance and scale-height. These reduced $E(B$--$V)_{\rm SFD}$ values are 0.13 (NGC~2323) and 0.02 (NGC~2539). Here we adopted H$=$125 pc (Bonifacio \etal 2000).

\renewcommand{\arraystretch}{1.4} 
\begin{table*}[!t]  
\begin{center}
\caption{The reddenings (Column 2), and the mean values of $\langle(U$--$B)_{Hyadas}\rangle$
of the Hyades reference line,  $\langle(U$--$B)_0\rangle$ values, for $\langle(B$--$V)_0\rangle$ (Cols.~3--5) as set by the iso-abundance lines for F-type stars of the two OCs. The $\delta(U$--$B)$, $\delta_{0.6}$, $[M/H]$ and $Z$ are given in Cols.~6--9 together their uncertainties.}
\label{tab:metallicities}
			\tiny
		\setlength{\tabcolsep}{0.3cm}
		\begin{tabular}{llccccccc}
			\hline
			Cluster&$E(B$--$V)$ &$\langle(B$--$V)_0\rangle$ & $\langle(U$--$B)_{H}\rangle$ &$\langle(U$--$B)_0\rangle$&$\delta(U$--$B)$& $\delta_{0.6}$&[M/H] &$Z$\\
			\hline
			NGC~2323&0.23$\pm$0.04  &0.48 &+0.01 &$-$0.03 &0.04 &0.044$\pm0.03$&-0.10$\pm0.11$  &0.012$\pm0.003$ \\
			NGC~2539&0.02$\pm$0.06  &0.51 &0.06 &$-$0.03 &0.03 &0.031$\pm0.03$&-0.31$\pm0.12$  &0.007$\pm0.003$ \\
			\hline
		\end{tabular}
	\end{center} 
\end{table*}

\subsection{Distance moduli, distances and ages}

For the determination of distances and ages of the two OCs, we have used the  PARSEC isochrones of Bressan \etal (2012) for $Y$ values which correspond to $Z$. Here, Y$=$0.2485+1.78Z. The PARSEC isochrones of Bressan \etal (2012) are over-plotted in four CMD's: $V,(B$--$V)$, $V,(V$--$I)$, $V,(R$--$I)$, $G,(G_{BP}$--$G_{RP})$ (Figs.~7--8 and Figs.~10--11). The $E(V$--$I)$, $E(R$--$I$) and $E(G_{BP}$--$G_{RP}$) colour excesses are converted from the relations $E(V$--$I)=1.25E(B$--$V)$, $E(R$--$I)=0.69E(B$--$V)$ (Dean \etal 1978, Mathis 1990 and Strai\c{z}ys 1995) and $E(G_{BP}$--$G_{RP})=0.775E(B$--$V)$ (Bragaglia \etal 2018). A visual extinction of $A_{\rm V} = 3.1\times E(B$--$V)$ is applied to the absolute visual magnitudes of the isochrones.

The PARSEC isochrones are first shifted both vertically and horizontally on the CMDs according to the interstellar reddening values of $E(B$--$V)$, $E(V$--$I)$, $E(R$--$I)$, and $E(G_{BP}$--$G_{RP}$). Then the PARSEC isochrones have been shifted vertically to obtain the best fit to the observed main sequence, as well as the RC sequence. This vertical shift gives the (true) distance modulus, $DM =(V_{0}$--$M_{\rm V})$. The distance moduli ($V_{0}$--$M_{\rm V}$) and distances, d(kpc), for the two OCs are presented in Cols.~2--3 of Tables~7--8.

For the determination of the ages (A,~log(\rm A)) of these OCs, the isochrones of PARSEC, selected according to their $Z$ values, have been shifted both vertically and horizontally in the CMD's with the expression $M_{V} + 3.1E(B$--$V) + DM$, for the vertical displacement and $C_0(\lambda_1-\lambda_2)+E(\lambda_1-\lambda_2)$, for the horizontal, where $\lambda$ denotes the wavelengths of the various passbands. Here C$_{0}$ means de-reddened colour index. Then the age of the isochrone is varied until a satisfactory fit to the data has been obtained through the observed main-sequence (MS), turn-off (TO), sub-giant (SG) and RG/RC sequences. The derived ages are given in Cols.~4--5 of of Tables~7--8. 

To appreciate the uncertainties of the distance moduli and ages, additional PARSEC isochrones have usually been over-plotted in each CMD. The best fit is shown by red solid line, whereas, the uncertainties are represented by gray solid lines. The photometric uncertainties of colour indices are indicated in bottom left of CC and CMD plots.

For comparisons with the literature, we have taken into consideration mostly those physical parameters given by our $(B$--$V)$ colour indices, because the astrophysical parameters of these OCs are mostly given, and best represented, in terms of the CMD: $V$,$(B$--$V)$ (Table 11).

\section{Results}
\renewcommand{\arraystretch}{1.7} 
\begin{table}[!t]  
	\caption{The derived fundamental astrophysical parameters of NGC~2323 four colour indices. Its reddening, metal and heavy element abundances are as the following. $E(B$--$V)=0.23\pm0.04$, $[M/H]=-0.10\pm0.11$, $Z=0.012\pm0.003$.}
	\label{tab:NGC2323parameters} 
	\setlength{\tabcolsep}{0.38cm}
	{\tiny 
		\begin{tabular}{lcccc}
			\hline
			Colour &$(V_{0}$--$M_{V})$ &  d~(kpc) & log(A)  & A~(Gyr) \\
			\hline
			$(B$--$V)$ &10.00$\pm$0.10 &1.00$\pm$0.05 &8.30$\pm$0.10 &0.20$\pm$0.05 \\
			$(V$--$I)$ &09.90$\pm$0.15 &0.96$\pm$0.07 &8.30$\pm$0.15 &0.20$\pm$0.08 \\
			$(R$--$I)$ &09.90$\pm$0.15 &0.96$\pm$0.07 &8.30$\pm$0.15 &0.20$\pm$0.08 \\
			$(G_{BP}$--$G_{RP})$ &09.80$\pm$0.10 &0.91$\pm$0.04 &8.35$\pm$0.10 &0.20$\pm$0.06 \\
			\hline
		\end{tabular} 
	}
\end{table}

\subsection{NGC~2323}
From 11 early type stars with $(U$--$B)<0.0$ in Fig.~6,  the reddening $E(B$--$V)=0.23\pm0.04$ of NGC~2323 is determined. As is seen from Table 11, this reddening value is quite consistent with the literature values, $E(B$--$V)=0.20$--0.28, within the uncertainties. Our reddening value and the literature values are larger than $E(B$--$V)_{\rm SFD}=0.13$. NGC~2323's $\delta(U$--$B)_{0.6}$ gives the photometric abundance of ($[M/H], Z) = (-0.10, 0.012$) from the CC diagram (Fig.~6).

\begin{figure}[!t]\label{ngc2323ubv.jpg}
	\includegraphics[width=0.95\columnwidth]{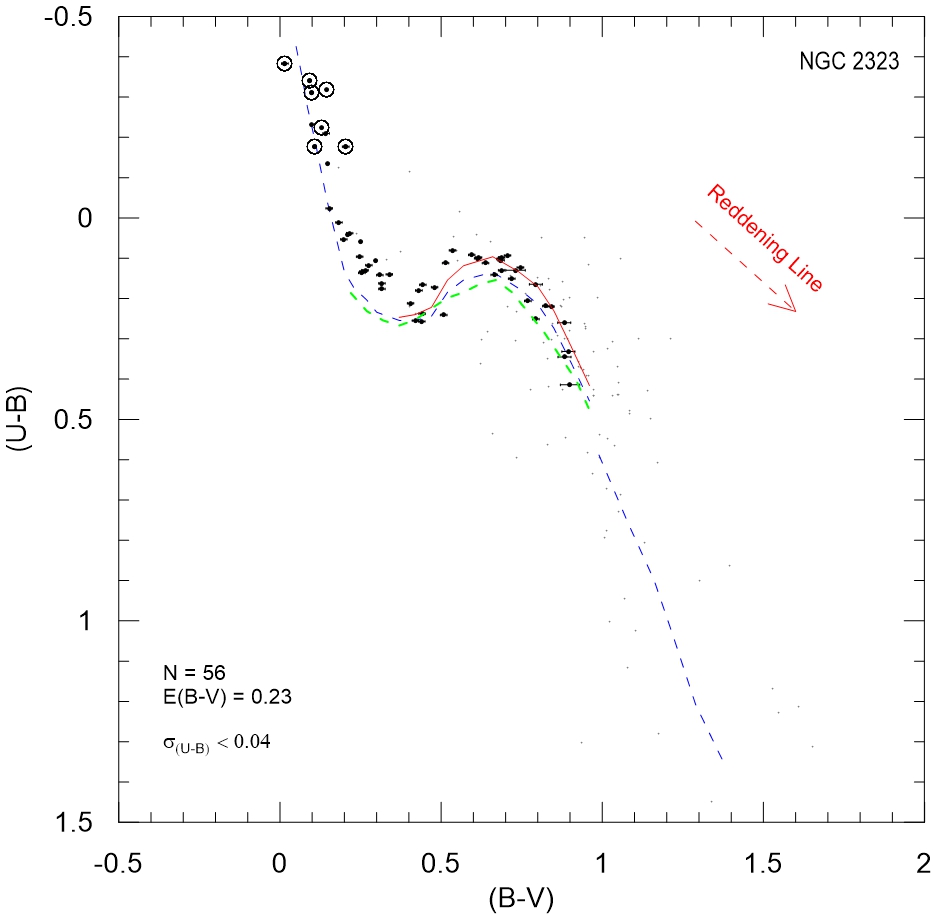}
	\caption{The reddened $(U$--$B, B$--$V)$ diagram for the likely members of NGC~2323.
		The blue dashed line shows the SK82 relation for both the main sequence (upper part) and
		the red giants (lower part).  The green dashed line denotes the Hyades main-sequence.
		The fitted iso-metallicity line is indicated by the solid red curve.  Grey dots indicate
		the non-members.  The reddening vector is the dashed red arrow, and big circles indicate
		the seven definite members (See Table~10).}
\end{figure}

\begin{figure}[!t]\label{ngc2323cmd1.jpg}
	\includegraphics[width=0.92\columnwidth]{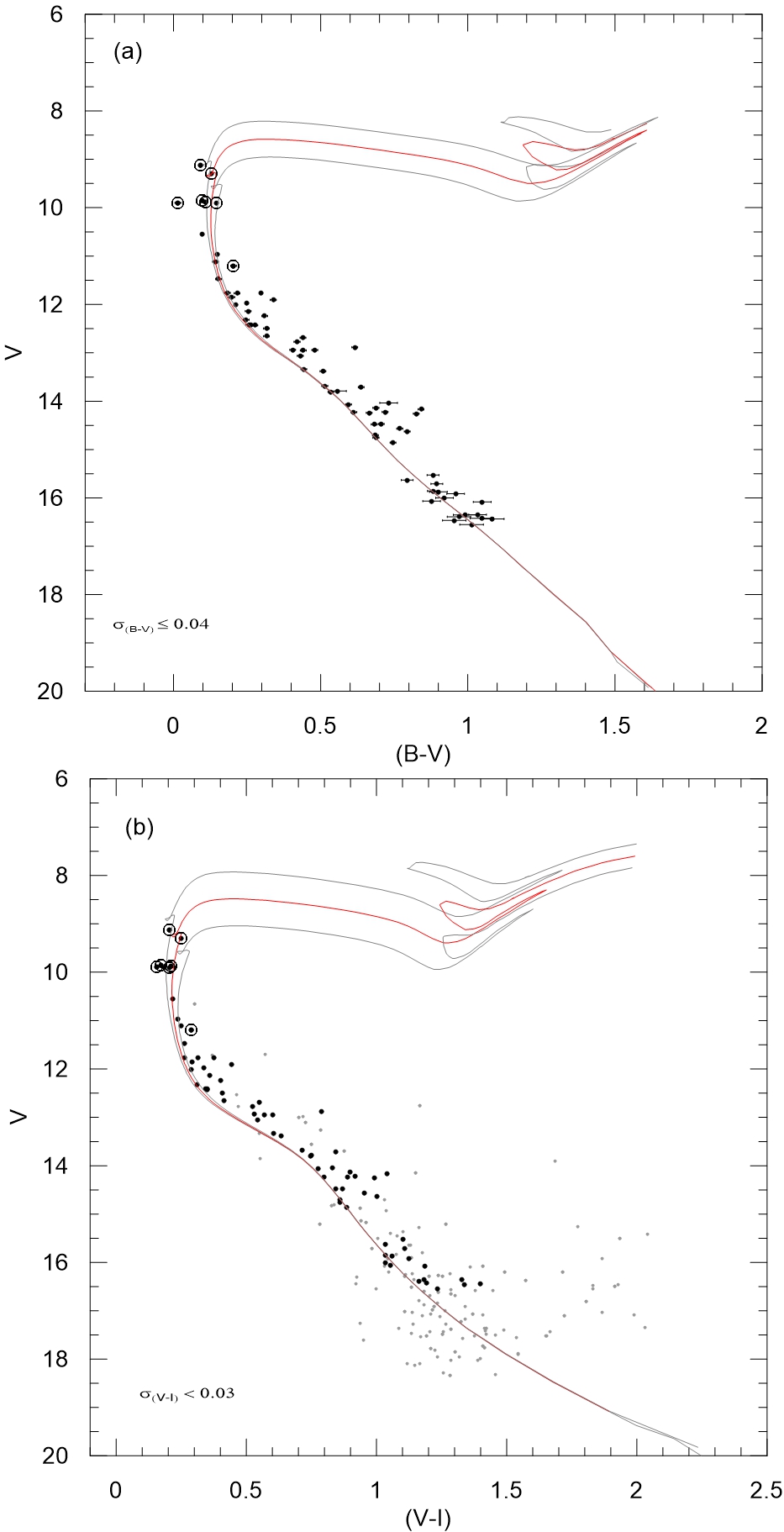}
	\caption{For NGC~2323, CMDs of $(V,B$--$V)$ (panel a) and $(V,V$--$I)$
		(panel b).  Red curves show the PARSEC isochrones interpolated to $Z=+0.012$.  
		Solid grey isochrones have been drawn to provide a means for appreciating the uncertainties of the ages.
		Filled and grey dots indicate the members and non-members, respectively.}
\end{figure}

\begin{figure}[!t]\label{ngc2323cmd2.jpg}
	\includegraphics[width=0.9\columnwidth]{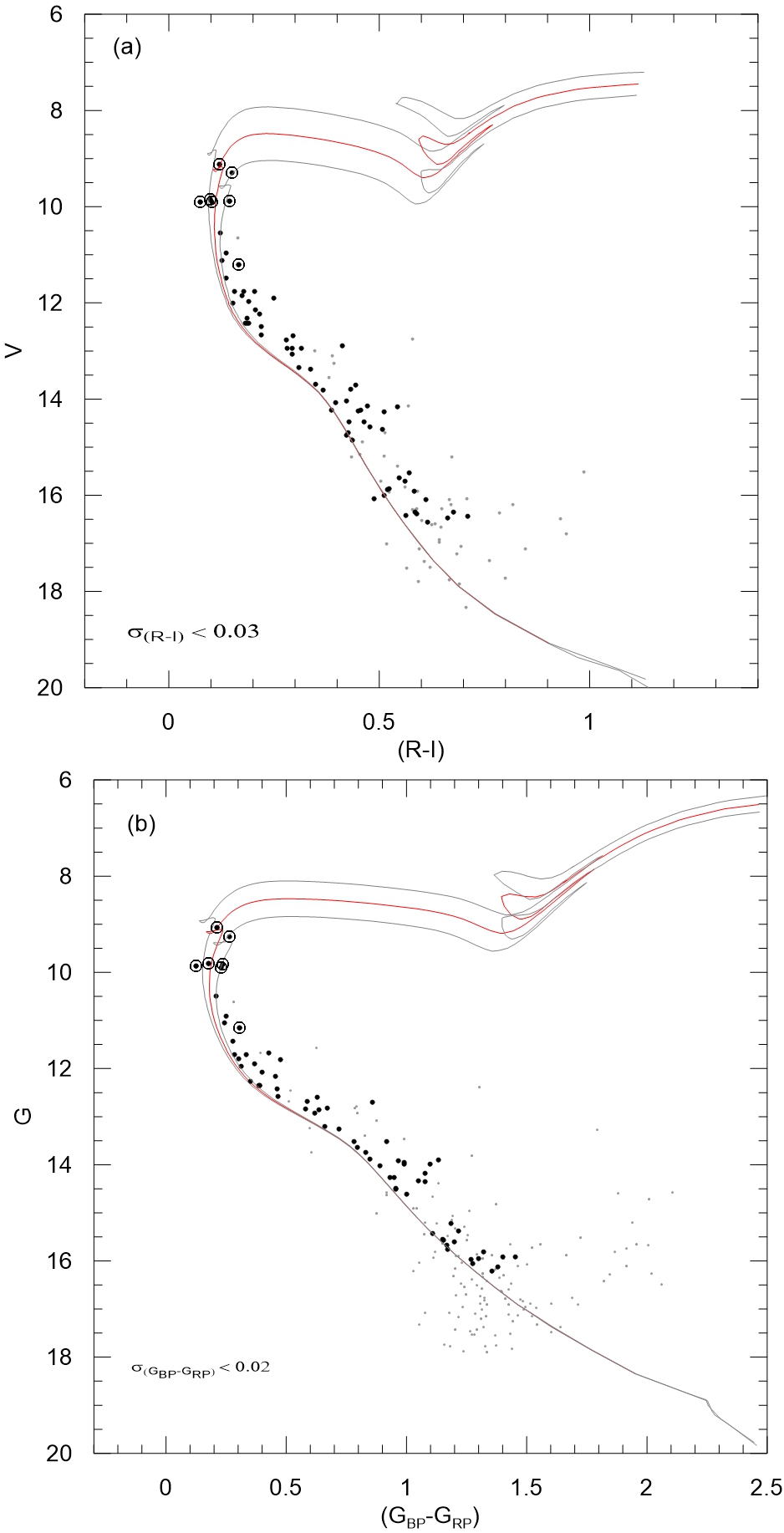}
	\caption{For NGC~2323, CMDs of $(V,R$--$I)$ (panel a) and $(G,G_{BP}$--$G_{RP})$ (panel b).
		The symbols are the same as Fig.~7.}
\end{figure}

From the CMDs for the colour indices $(B$--$V)$, $(V$--$I)$, $(R$--$I)$ and $(G_{BP}$--$G_{RP}$) (Figs.~7--8), values for the distance moduli, distances, and ages together with their uncertainties for NGC~2323 are presented in Table~7. The 200 Myr PARSEC isochrone with $Z=0.012$ fits well the main sequence for all colour indices. No evolved stars are seen in the CMDs of NGC~2323. The PARSEC isochrones fit well the evolutionary sequences, as is seen from the CMDs. We derived the distance modulus and distance as ($V_{0}$--$M_{\rm V}$, ~d(kpc))=(10.00$\pm$0.10, 1.00$\pm$0.05), by fitting the PARSEC isochrones to the $(V, (B$--$V))$ diagram (Fig.~7(a)). The differences with the literature are at a level of $\Delta (V_{0}$--$M_{\rm V})=0$--$0.46$ mag and $\Delta d=0$--$0.11$ kpc.  Our age value is somewhat older than the age range of 100--140 Myr in the literature (Table 11). This is to be expected since we used $Z=0.012$ instead of the solar metal abundance. The distance modulus and distance of this OC are in agreement with the literature values (Table 11). Note that some authors of the literature do not give their uncertainties.

Our CCD $UBV(RI)_{KC}$ photometric data of NGC 2323 contain seven definite early-type members, according to the SIMBAD data-base. The GaiaDR2 parallaxes/distances plus UBV photometry of these seven definite members are listed in the top rows of Table~10. These are also marked as big circles in the CC and CMDs (Fig.~6 and Figs.~7--8). Our photometric distances of $(B$--$V)$, $(V$--$I)$, $(R$--$I)$, and $(G_{BP}$--$G_{RP}$) provide good agreement with the Gaia DR2 distances (Col.~7, Table~10) within their uncertainties.

\subsection{NGC~2539}
A reddening value, $E(B$--$V)=0.02\pm0.06$ for NGC~2539 (Fig.~9) is derived. Within the uncertainties, this value is in good coherent with the ones of the literature (Table~11). The reduced $E(B$--$V)_{\rm SFD}=0.02$ value for this OC is in good agreement with our reddening value. NGC~2539's $\delta(U$--$B)_{0.6}$ gives the photometric abundance of ($[M/H], Z) = (-0.31, 0.007$) from the CC diagram (Fig.~9).

\renewcommand{\arraystretch}{1.4} 
\begin{table}[!t]  
	\caption{The derived fundamental astrophysical parameters of NGC~2539 four colour indices. Its reddening, metal and heavy element abundances are as the following. $E(B$--$V)=0.02\pm0.06$, $[M/H]=-0.31\pm0.12$, $Z=0.007\pm0.003$.}
		\label{tab:NGC2539parameters} 
	{\tiny \tiny
		\setlength{\tabcolsep}{0.39cm}
		\begin{tabular}{lcccc}
			\hline
			Colour &$(V_{0}$--$M_{V})$ &  d~(kpc) & log(A)  & A~(Gyr) \\
			\hline
			$(B$--$V)$ &10.00$\pm$0.04 &1.00$\pm$0.02&8.95$\pm$0.05 &0.89$\pm$0.11 \\
			$(V$--$I)$ &10.00$\pm$0.10 &1.00$\pm$0.05&8.95$\pm$0.10 &0.89$\pm$0.23 \\
			$(R$--$I)$ &10.00$\pm$0.10 &1.00$\pm$0.05&8.95$\pm$0.10 &0.89$\pm$0.23 \\
			$(G_{BP}$--$G_{RP})$&10.10$\pm$0.10&1.00$\pm$0.05&8.95$\pm$0.10 &0.89$\pm$0.23\\
			\hline
		\end{tabular} 
	}
\end{table}   

\begin{figure}[!t]\label{ngc2539ubv.jpg}
	\includegraphics[width=0.95\columnwidth]{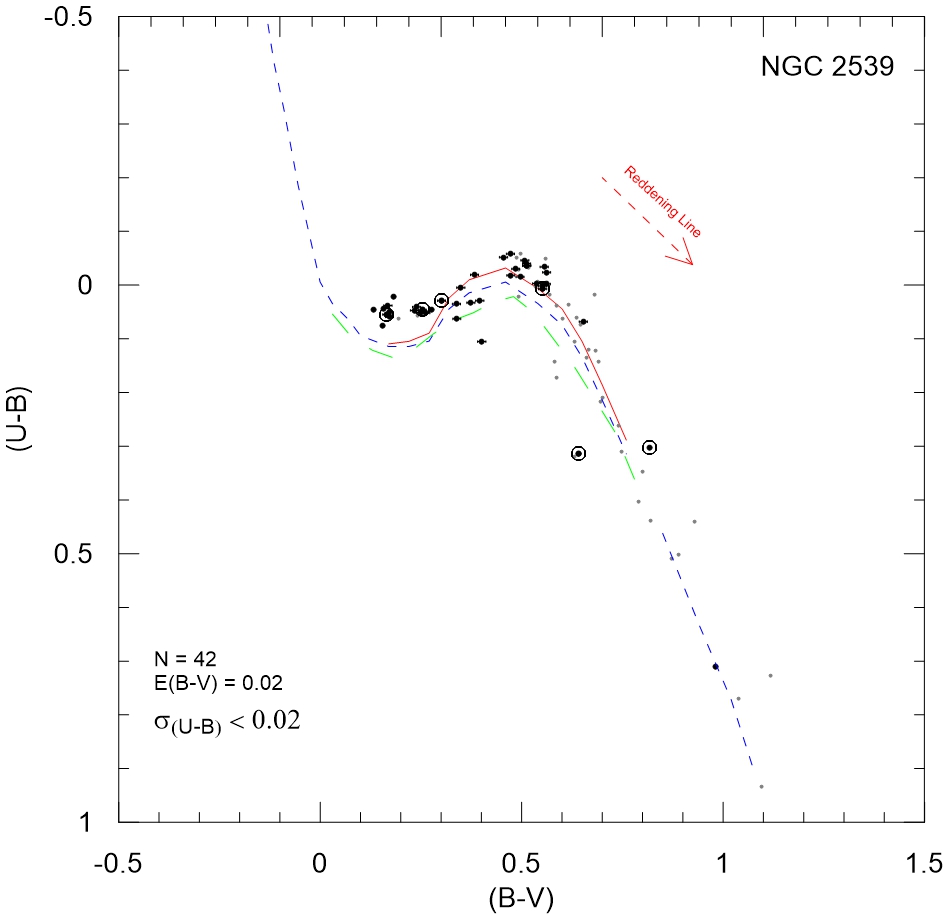}
	\caption{The reddened $(U$--$B, B$--$V)$ diagram for NGC~2539. 
		The six definite members are shown as circles.  
		The meanings of the symbols are the same as Fig.~6}
\end{figure}

\begin{figure}[!t]\label{ngc2539cdm1.jpg}
	\includegraphics[width=1.0\columnwidth]{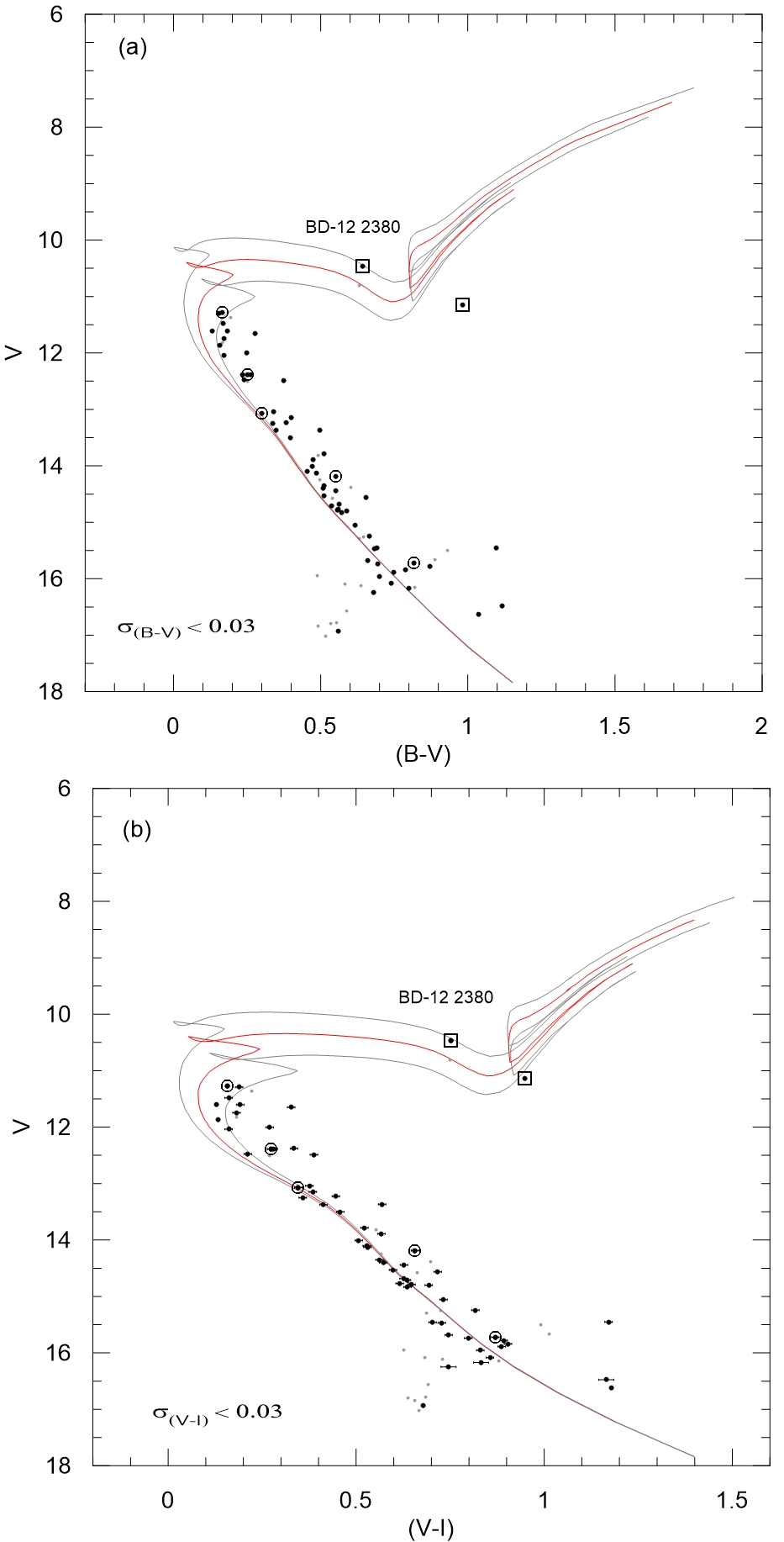}
	\caption{For NGC~2539, CMDs of $(V,B$--$V)$ (panel a) and $(V,V$--$I)$  (panel b).
		Red curves show the PARSEC isochrones interpolated to $Z=+0.007$.  Solid grey isochrones have been
		drawn to provide a means for appreciating the uncertainties of the ages. Big open circles mark
		six definite members.  Note that the two giant members (square dots) with $(V,B$--$V)=$(10.462,~0.642) (BD-12 2380) and $(V,B$--$V)=$(11.144,~0.982).}
\end{figure}

\begin{figure}[!b]\label{ngc2539cdm2.jpg}
	\includegraphics[width=1.0\columnwidth]{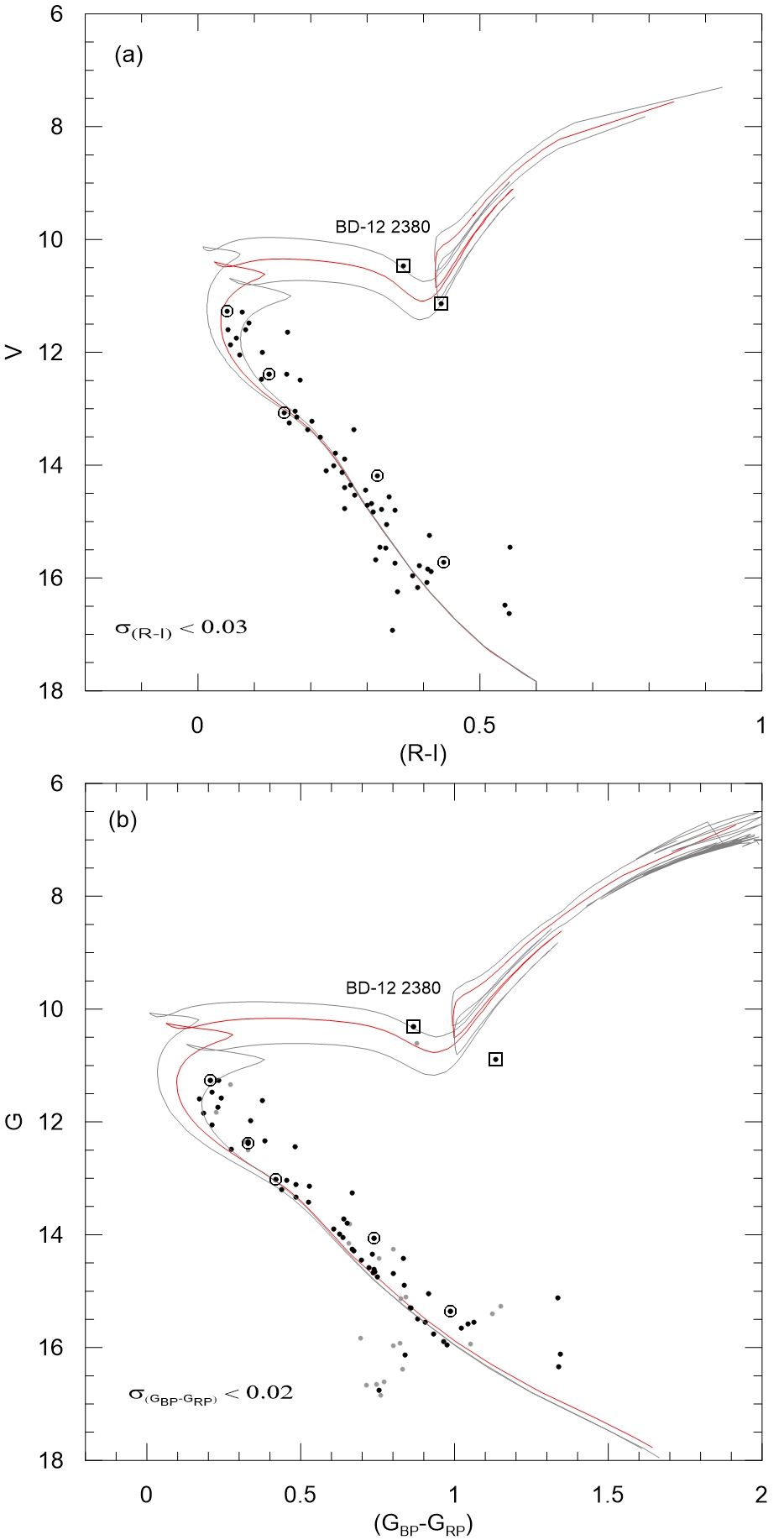}	
	\caption{For NGC~2539, CMDs of $(V,R$--$I)$ (panel a) and $(G,G_{BP}$--$G_{RP})$ (panel b). 
		The meanings of the symbols are the same as Fig.~10.}
\end{figure}

The CMDs for the colour indices $(B$--$V)$, $(V$--$I)$, $(R$--$I)$, $(G_{BP}$--$G_{RP}$) of NGC~2539 are presented in panels~(a)--(b) of Figs.~10--11. The isochrone with $Z=0.007$ fits well the main sequence for all colour indices. The derived fundamental parameters together with their uncertainties from these CMDs have been given in Table~8. For the $(B$--$V)$ colour, our values $(V_{0}-M_{\rm V})=10.00\pm0.04$ and $d$(kpc)$=1.00\pm0.02$ are in good concordance with the ones of the literature within the uncertainties (Table~11). The differences with the literature for NGC~2539 are at a level of $\Delta (V_{0}$--$M_{\rm V})=0.10-0.50$ mag and $\Delta d=0.05-0.29$ kpc.
Our age value 0.89$\pm$0.11 Gyr (890 Myr) for NGC~2539 seems to be slightly older than the age range of 540--640 Myr of the literature (Table 11).
 
No RC/RG candidates are detected in the CMDs of NGC~2323 because of its younger age ($A = 0.20$ Gyr). Note that the possible RC/RG candidates in the CMDs of NGC~2539 (square dots,~Figs.~10--11). One of them is named BD-12 2380, according to the SIMBAD data-base. One can expect to detect this possible RC/RG star in the central part of NGC~ 2539, depending on its age ($A = 0.89$ Gyr).  For the this giant candidate, from the $V,(B$--$V)$  and $V,(V$--$I)$ CM diagrams (square dot, panels~(a)--(b) of Fig.~10), we have utilised the distance criteria for its  membership as a initial test. In the case of the possible RG candidates, the RG candidates will not necessarily give the right distances to these clusters since their apparent magnitudes will depend on their position along the RG branch.

For the case of a possible RC candidate, its distances have been estimated from the mean absolute magnitudes of $\langle M_{\rm V} \rangle=+0.60\pm0.10$, (Twarog \etal 1997) and $\langle M_{\rm I} \rangle=-0.22\pm0.03$ (Groenwegen 2008), and listed in Table~9 together with its equatorial coordinates plus its magnitudes of $I$ and $V$. For the estimates, the total absorption relations of ${A_{\rm V}} = 3.1~E(B$--$V)$  and $A_{\rm I} = 1.98~E(B$--$V)$ (Gim \etal 1998), are included, with the $E(B$--$V)=0.02$ value of NGC~2539. The magnitudes of $I$ and $V$ in Columns~3--4 of Table~9 are used for estimating the $d_{\rm I}$ and $d_{\rm V}$ distances. These distances are estimated as $d_{\rm I}=0.95\pm0.05$ kpc and $d_{\rm V}=0.91\pm0.09$ kpc. The giant candidates within 1-$\sigma$ agreement of the uncertainties of $d_{\rm I}$ and $d_{\rm V}$ from the main-sequence fitting (Table~8) have been assigned as members, ``M'', otherwise as non-members ``NM'' (Columns~7--8 of Table~9). The giant star, BD-12 2380 which fulfills both the $d_{\rm I}$ and $d_{\rm V}$ distances within its uncertainties is a definite member of NGC~2539. Its GaiaDR2 distance, $d=1.20\pm0.07$ kpc ($\varpi = 0.833\pm0.047$) mas is close to the photometric distances of Tables 8--9.

The other giant candidate does not seem to be RC candidate due to its location, $(V, B$--$V)=$(11.144,~0.982) in the CMDs (square dot,~Figs.10--11). Its GaiaDR2 distance $d=1370\pm80$ pc ($\varpi = 0.732\pm0.041$ mas) is not close to the photometric distance, 1000 pc of Table~8. 

From TO and RC/RG sequences on the CMDs of NGC~2539, for BD-12 2380 RC/RG candiate, we have determined its  morphological age index (MAI). For this, $\delta V$ and $\delta 1$ indices given by Phelps \etal (1994) are determined within its CMDs. Here,  $\delta V$ is the magnitude difference between the TO and RC stars, $\delta V = V_{TO} - V_{RC}$, and $\delta 1$ is the difference in the colour indices between the bluest point on the main sequence at the luminosity of the TO and the colour at the base of RG branch one magnitude brighter than the TO luminosity,  $\delta 1=(B$--$V)_{\rm TO}-(B$--$V)_{\rm RG}$. We measured the values of $\delta 1$ from its RC candidate, and then converted into $\delta V$ by means of the equation of $\delta V=3.77-3.75\delta 1$ of Phelps \etal (1994). The photometric values obtained are listed at bottom of Table~9. With the aid of its photometric metal abundance, its morphological age is determined from the equation, $log A=0.04\delta V^2+0.34\delta V+0.07[Fe/H]+8.76$ of Salaris \etal (2004). The estimated MAI age is listed in the bottom line of Table~9 (Cols.~7--8) together with its uncertainty. The age difference between the MAI and isochrone techniques is $\Delta log A = 0.09$ ($\Delta A = 0.20$ Gyr). These quite small differences indicate good consistency between the techniques.

The GaiaDR2 parallaxes/distances plus $UBV$ photometry of six definite members which belong to NGC~2539, according to SIMBAD-database are listed in the bottom rows of Table~10. Differences fall in the range of $\Delta d$ = 200--456 pc between the 1000~pc value of all colours (Table~8) and the ones of GaiaDR2 (Col.~7; Table~10). In the sense their distances of these six definite members are not close to the photometric distances of NGC~2539 of four colour indices within the uncertainties.

\renewcommand{\arraystretch}{1.5} 
\begin{table}[!t]  
	\caption{First two rows show the membership test according to the distance criterion 
		for the possible Red Clump (RC) member of NGC~2539. RA, DEC (Cols.1-2), $VI$ photometry (Cols.~3-4), Distances for I and V magnitudes (Cols.~5-6), test distances (Cols.~7-8).
		``M'' in Cols.~7--8 means member.  The morphological age
		indices (MAI) of NGC~2539 are given at the bottom. 
		``TO'' and ``RC'' mean Turn-Off and Red Clump stars,
		respectively.  The morphological age values, $log A$ and
		$A$ (Gyr) are listed in Cols.~7--8.}
	\setlength{\tabcolsep}{0.14cm}
	{\tiny 
		\begin{tabular}{lcccccccc}
			\hline
			RA  &  DEC  &   I  &   V  & $d_{I}$ (kpc)  & $d_{V}$ (kpc)  & $d_{I}$(test) & $d_{V}$(test) \\
			&       &      &        &         &           &   &   \\
			\hline
			122.71 & -12.88  &  9.71   & 10.462 & 0.95$\pm$0.05 & ~0.91$\pm$0.09 &  M   & M\\
			122.64 & -12.86  & 10.20   & 11.144 & 0.96$\pm$0.06 & ~1.23$\pm$0.14 &  M   & NM\\
				&       &      &        &         &    &   &   \\
			\hline
			V$_{TO}$ & V$_{RC}$ & (B-V)$_{TO}$ & (B-V)$_{RC}$ & $\delta V$ & $\delta 1$ & log A& A~(Gyr)\\
			\hline
			11.40  &  10.70   &  0.08  &  0.89 &  0.73 &  0.81 & 9.04$\pm$0.05 & 1.09$\pm$0.27 \\
			\hline				
		\end{tabular} 
	}
\end{table}   

\renewcommand{\arraystretch}{1.4} 
\begin{table}[!t]  
	\caption{Our $UBV$ photometry and Gaia DR2 parallaxes for the definite members of
		NGC~2323 (seven) and NGC~2539 (six), according to the SIMBAD data-base.
		Star designations (Cols.~1--2), $V$, $(U$--$B)$, $(B$--$V)$ (Cols.~3--5).
		GaiaDR2 parallaxes (mas) and their converted distances (pc) together
		with the uncertainties  (Cols.~6--7).}
	{\tiny 
		\setlength{\tabcolsep}{0.22cm}
		\begin{tabular}{lllllrr}
			\hline
			Star-ID&Star No&V &(U-B) &(B-V) &$\varpi$ $\pm$ $\sigma_{\varpi}$ &d $\pm$ $\sigma_{d}$ \\
			\hline
			NGC~2323& & & & & &   \\
			\hline
			228&HD 52965 &9.13 &-0.34&0.09&1.004$\pm$0.042&996$\pm$42 \\
			163&BD-08 1703&9.30 &-0.22&0.13&1.170$\pm$0.045&855$\pm$33 \\
			256&BD-08 1700& 9.85 &-0.31&0.10&0.914$\pm$0.047&1094$\pm$56 \\
			402&BD-08 1695& 9.88 &-0.18&0.11&0.967$\pm$0.046&1034$\pm$49 \\
			359&BD-08 1696& 9.90 &-0.38&0.02&1.008$\pm$0.047&992$\pm$46 \\
			70&BD-08 1708 & 9.90 &-0.32&0.14&0.951$\pm$0.060&1052$\pm$66 \\
			105&HD 52980B&11.20 &-0.18&0.20&1.023$\pm$0.041&978$\pm$39  \\
			\hline
			NGC~2539& & & & & &  \\
			\hline
			37 & BD-12 2380   &10.46&0.31&0.64& 0.833$\pm$0.047& 1200$\pm$70  \\
			168& BD-12 2373   &11.28&0.06&0.16& 0.713$\pm$0.063&1403$\pm$120 \\
			234&V0691Pup      &12.39&0.05&0.25&0.796$\pm$0.048 &1256$\pm$80   \\
			122&V0693Pup      &13.07&0.03&0.30&0.782$\pm$0.038 &1279$\pm$60 \\
			98 &V0694Pup      &14.18&0.01&0.55&0.687$\pm$0.034 &1456$\pm$70  \\
			79 &V0695Pup      &15.72&0.30&0.82&0.730$\pm$0.050 & 1370$\pm$90 \\
			\hline
		\end{tabular}
	}
\end{table}


\renewcommand{\arraystretch}{1.6} 
\begin{table*}[!t]  
	\caption{Comparison with the literature for our sample of OCs.  The comparison is given
		here for our $(B$--$V)$ results.}	
	{\tiny \tiny
		\setlength{\tabcolsep}{0.33cm}
		\begin{tabular}{lllllllllll}
			\hline 
			Cluster &E(B-V) &$(V_{0}$--$M_{V})$&    d(kpc) & Age (Myr) &log A &    Z & Isochrone & Photometry &  Ref.\\
			\hline 
			NGC 2323  &0.23$\pm$0.04 &10.00$\pm$0.10 &1.00$\pm$0.05&200$\pm$50&8.30 &0.012 
			& Bressan et al.(2012) &CCD UBVRI &This work  \\
			
			&0.23$\pm$0.06 &10.00$\pm$0.15&1.00$\pm$0.07&140$\pm$20   &8.15  &solar&Yi et al. (2001); &UBV; CFHT &1 \\
			
			& - & -& - &115$\pm$20   &8.06  &solar&Bressan et al.(2012)&UBV; CFHT  &1  \\
			
			&0.23 & -  &0.895 &100   &8.00&-    &-   &- &2  \\
			&0.25 &9.85&1.00 &140 &8.15&0.019&Girardi et al.(2000)&Various broadband photometries  &3  \\
			&0.28 &10.46&1.107$\pm$0.07&-&8.30&0.019&Girardi et al.(2002)&2MASS JH${K_{s}}$&4  \\
			&0.20 &10.50&0.95 &100&8.00&0.020&Bertelli et al.(1994)&CCD UBV  &5  \\
			&0.22 &10.00$\pm$0.17 &1.00$\pm$0.08 &130 & -  &0.020&Ventura et al.(1998) &BV with CFHT12Mosaic camera&6  \\
			
			&0.25$\pm$0.05 &9.86 &0.94&100$\pm$20 &8.00 &0.020&Bertelli et al.(1994)  &Photoelectric UBV &7 \\
			
			&0.257  & - & - & - &  - & - &- &Stromgren&8 \\
			&-  &- &0.997$\pm$0.057& - & -  &-&  -     &GaiaDR2 astrometry&9 \\
			
			& & &    &    &    &     &     &    &   \\
			\hline         
			NGC 2539 &0.02$\pm$0.06 &10.00$\pm$0.04&1.00$\pm$0.02&890$\pm$110 &8.95  &0.007&Bressan et al.(2012)   &CCD UBVRI                       &This work \\
			&0.06$\pm$0.03 &10.20$\pm$0.10&1.10$\pm$0.05&630          &8.80  &0.019&Girardi et al.(2000)   &CCD UBVI                        &10  \\
			&0.06          &10.42         &       1.21  &630          &8.80  &0.020&Schaller et al.(1992)   &photoelectric UBV               &11  \\
			&0.08$\pm$0.02 &10.10$\pm$0.30&1.05$\pm$0.15&540          &8.73  &0.030&Hejlesen (1980)        &photoelectric UBV               &12  \\
			&0.08$\pm$0.02 &09.80$\pm$0.50&0.91$\pm$0.21&640$\pm$80   &8.81  &0.016&Mermilliod (1981)      &CCD DDO, CMT$_{1}$T$_{2}$, UBV  &13 \\
			&0.10$\pm$0.05 &10.50$\pm$0.50&1.29$\pm$0.29& -  &  -   &   - &     -  &photoelectric UBV  &14 \\
			&- &- &0.754$\pm$0.064& - & -  &-&  -     &GaiaDR2 astrometry                &9\\ 
			\hline
		\end{tabular}
	} \vspace{2mm}
	{\tiny \tiny
		\begin{list}{Table Notes.}
			\item (1):Cummings, J et al. (2016), AJ, 818, 84; (2):Paunzen \& Netopil (2014); (3): Frolov, V. N. et al. Ast. Lettr., (2012), 38,74; (4): Bukowiecki, L. et al. (2011),  Acta Astronomica, 61, 231;
			(5): Sharma, S. et.al. (2006), AJ, 132, 1669; (6): Kalirai, J.S. et.al. (2003), AJ, 126, 1402
			;(7):Claria, J.J. et.al. (1998), A\&ASS, 128, 131;(8) :Schneider, H., (1987), A\&ASS, 71, 531;  (9): Cantat-Gaudin, T. et.al. (2018), Astronomy and Astrophysics, 618, 93;  (10):Choo, K.J. et.al. (2003), A\&A, 399, 99; (11):Lapasset, E. et.al. (2000), A\&A, 361, 945
			;(12):Joshi\&Sagar, (1986), Bull.Astr. Soc. India, 14, 95; (13):Claria\&Lapasset (1986), ApJ, 302, 656; (14):Pesch (1961), ApJ, 134, 602.  
		\end{list}
	}
\end{table*}

\section{Discussions and Conclusions}

The reddenings of $E(B$--$V)=0.23\pm0.04$ of NGC~2323 and $E(B$--$V)=0.02\pm0.06$ of NGC~2539 are quite consistent with the ones of the literature (Table~11). The reddening differences with the literature  are at a level of 0.00--0.05 for  NGC~2323 and 0.04--0.08 for NGC~2539. 
For the $(B$--$V)$ colour, distance moduli and distances of NGC~2323 and NGC~2539 are ($V_{0}$--$M_{\rm V}$, $d$ (pc)) = (10.00$\pm$0.10, 1000$\pm$50) and ($V_{0}$--$M_{\rm V}$, $d$ (pc)) = (10.00$\pm$0.04, 1000$\pm$20), respectively. The differences with the literature, derived using the $(B$--$V)$ colour, are at a level of $\Delta (V_{0}$--$M_{\rm V}) = 0.00-0.46$ mag and $\Delta d$ (kpc)$ = 0.00$--0.11 for NGC 2323.
These  differences for NGC~2539 are at a level of $\Delta (V_{0}$--$M_{\rm V}) = 0.10-0.50$ mag and $\Delta d$ (kpc) $= 0.05$--0.29. In the sense our detections are in good concordance with the values of the literature within the uncertainties (Table 11). For four colour indices (Tables 7--8), the values of ($V_{0}$--$M_{\rm V}$, $d$ (kpc)) are 9.80--10.00 and 0.91--1.00 kpc for NGC~2323, and 10.00 and 1.00 kpc for NGC~2539, respectively, which are in agreement with each other and the literature values (Table 11).  

The median GaiaDR2 parallax ($\varpi$=0.998$\pm$0.136 mas) of NGC 2323 (74 likely members) provides $d=1000\pm140$ pc. This value is also similar to the photometric ones ($d = $910--1000 pc) (Table~7), and also in good agreement with the distances of the literature (Table 11). For NGC~2539 (N=67 likely members), the median GaiaDR2 parallax  $\varpi$=0.751$\pm$0.139 mas measures $d=1330\pm 250$ pc, which is reasonably close to our 1000 pc value (Table~8) within the uncertainties. For NGC~2539, the literature values are not as close to each other: $\varpi=0.91$ mas (d $=$ 1.10 kpc) and $\varpi=0.78$ mas (d $=$ 1.29 kpc). Our median parallaxes are similar to the values 0.997$\pm$0.057 mas (NGC~2323) and 0.754$\pm$0.064 mas (NGC~2539) of Cantat-Gaudin \etal (2018) (Table 11).

Our age value (200$\pm$50 Myr) of NGC~2323 is somewhat older than the age range of 100$-$140 Myr in the literature (Table~11). This occurs because we estimated $Z=0.012$ from the ultraviolet excess (Fig.6), instead of assuming the solar metal abundance. Likewise our estimated age, 890$\pm$110~Myr for NGC~2539 is somewhat older than the age range 540--640 Myr of the literature (Table 11). The works of Joshi \& Sagar (1986), Claria \& Lapasset (1986), Choo \etal (2003) and Lapasset (2000) use different isochrones (Col.~8 of Table~11), and the differences in ages with respect to our values result from their usage of solar/different $Z$ abundances with these isochrones. This work uses the $Z=0.007$ abundance, which has been obtained from the $\delta(U$--$B)_{0.6}$ value of Fig.9. The morphological age of NGC~2539 (the MAI method) of the one probable RC star (Table~9) has been determined as 1.09$\pm$0.27 Gyr. This MAI age is quite compatible with its isochrone age (0.89 Gyr; Table~8).
	
The giant star, BD-12 2380 appears to be a definite member of NGC~2539 (square dot,~Figs.~10--11) since the distances of $d_{I}$ =950$\pm$50 pc and $d_{V}$=910$\pm$90 pc for $VI$ filters (Table 9) are in harmony with the photometric distance 1000$\pm$20 pc (Table 8) within the uncertainties. However, its GaiaDR2 distance $d=1200\pm70$ pc ($\varpi = 0.833\pm0.047$) mas is not close to the photometric distance of NGC~2539. The other giant candidate's  GaiaDR2 distance    is $d=1370\pm80$ pc ($\varpi = 0.732\pm0.041$ mas) but is not close to the photometric distance of NGC~2539. In the sense this may be a field giant. For further confirmation of its membership, spectroscopic observations are needed.

\section*{Acknowledgments}
The observations of this publication were made at the National Astronomical Observatory, San Pedro M\'artir, Baja California, M\'exico, and we wish to thank the staff of the observatory for their assistance during these observations. We thank H.Cakmak for useful helps on the manuscript.  This research made use of the WEBDA open-cluster database of J.-C. Mermilliod; also the SIMBAD data-base. This work has been supported by the CONACyT (M\'exico) projects 33940, 45014, and 49434, and PAPIIT-UNAM (M\'exico) projects IN111500 and IN103014. This paper has made use of results from the European Space Agency (ESA) space mission Gaia, the data from which were processed by the Gaia Data Processing and Analysis Consortium (DPAC). Funding for the DPAC has been provided by national institutions, in particular the institutions participating in the Gaia Multilateral Agreement. The Gaia mission website is http://www.cosmos.esa.int/gaia.

\vspace{-1em}



\begin{theunbibliography}{} 
\vspace{-1.5em}
\bibitem{akk10} Akkaya {\.I}., Schuster W.~J., Michel R., Chavarr\'ia-K C., Moitinho A., V\'azquez R. and Karata\c{s} Y., 2010, Rew Mex, 46, 385 (A10)
\bibitem{ora15} Akkaya Oralhan, I., Karata\c{s}, Y., Schuster,W.J., Michel, R., and Chavarr\'ia C., 2015, \na, 34, 195 (A15)
\bibitem{bs80} Bahcall, J.N., and Soneira, R.M., 1980, \apj, 238, 17	
\bibitem{ber94} Bertelli, G., Bressan, A., Chiosi, C., Fagotto, F., and Nasi,E. 1994, \aaps, 106, 275

\bibitem{bon} Bonifacio, P., Monai, S., and Beers, T.~C. 2000, \aj, 120, 2065
\bibitem{bra18} Bragaglia, A., Fu, X.,  Mucciarelli, A., Andreuzzi, G., Donati, P. 2018, \aap, 619, 176
\bibitem{bre12} Bressan A., Marigo, P., Girardi L., Salasnich, B., Dal Cero, C., Rubele, S., Nanni, A., 2012, \mnras, 427, 127
\bibitem{buk11} Bukowiecki {\L}., Maciejewski, G., Konorski, P., Strobel, A., 2011, \actaa, 61, 231
\bibitem{cae15} Caetano, T.C., Dias, W.S., L{\'e}pine, Monteiro, H.S., Moitinho, A., Hickel, G.R., Oliveira, A.F., 2015, \na, 38, 31
\bibitem{can18} Cantat-Gaudin, T., Jordi, C., Vallenari, A., Bragaglia, A.,  Balaguer-Nunez, L., Soubiran, C., Bossini, D., Moitinho, A., Castro-Ginard, A., Krone-Martins, A., and 3 coauthors., 2018, \aap, 618, 93
\bibitem{car89} Cardelli, J.A., Clayton G.C., Mathis, J.S., 1989, \apj, 345, 245
\bibitem{cho03} Choo, K. J., Kim, S.-L., Yoon, T. S., Chun, M.-Y., Sung, H., Park, B.-G., Ann, H. B., Lee, M. G., Jeon, Y.-B., Yuk, I.-S., 2003,  \aap, 399, 99	
\bibitem{cla86} Claria, J. J., Lapasset, E., 1986, \apj, 302, 656
\bibitem{cla98} Claria, J. J., Piatti, A. E., Lapasset, E., 1998, \aaps, 128, 131	
\bibitem{cum16} Cummings, J.D., Kalirai, J.S., Tremblay, P.-E., Ramirez-Ruiz, E., 2016, \apj, 818, 84
\bibitem{cut03} Cutri, R., \etal, 2003, $2MASS$ All-Sky Catalogue of Point Sources, CDS/ADC Electronic Catalogues 224
\bibitem{cut13} Cutri, R.M. \etal, 2013, yCat, 2328, 0C.

\bibitem{dea78} Dean, J.F., Warren, P.R., Cousins, A.W.J., 1978, \mnras, 183, 569

\bibitem{dia02} Dias, W.S., Alessi, B.S., Moitinho, A., L{\'e}pine J.R.D., 2002, \aap, 389, 871
\bibitem{fro95} Frolov, V. N., Ananjevskaja, Yu. K., Polyakov, E. V., 2012, AstL, 38, 74F

\bibitem{gaia1} Gaia Collaboration:  Brown,A.G.A, Vallenari, A., Prusti, T., de Bruijne, J.H.J.,  \etal, 2016, \aap, 595, A2 
\bibitem{gaia3} Gaia Collaboration:  Brown, A. G. A., Vallenari, A., Prusti, T. \etal, 2018, \aap, 616, 1G 
\bibitem{gaia5} Gaia Collaboration:  Lindegren, L., Hernandez, J., Bombrun, A., Klioner, S. \etal, 2018, \aap, 616, A2 
\bibitem{gim98} Gim, M., Vandenberg, D.A.,  Stetson, P., Hesser, J., Zurek, D.R., 1998, \pasp, 110, 1318 
\bibitem{gir00} Girardi, L., Bressan, A., Bertelli, G., Chiosi, C., 2000, \aaps, 141, 371
\bibitem{gir02} Girardi, L., Bertelli, G., Bressan, A., Chiosi, C., Groenewegen, M. A. T., Marigo, P., Salasnich, B., Weiss, A., 2002, \aap, 391, 195

\bibitem{gro08}Groenewegen, M.A.T., 2008, \aap, 488, 935 
\bibitem{hej80} Hejlesen, P. M., 1980,  \aaps, 39, 347
\bibitem{jos86} Joshi, U. C., Sagar, R., 2002, BASI, 14, 95	
\bibitem{joh53} Johnson, H.L., Morgan, W.W., 1953, \apj, 117, 313

\bibitem{kal03} Kalirai, J.S., Fahlman G.G., Richer, H.B., Ventura, P., 2003, \aj, 126, 1402	
\bibitem{kar06} Karata\c{s}, Y., Schuster W.J., 2006, \mnras, 371, 1793
\bibitem{kha13} Kharchenko, N.V., Piskunov, A.E., Schilbach, E., R{\"o}ser, S., Scholz, R.D., 2013, \aap, 558, 53
\bibitem{lan09} Landolt, A.U., 2009, \aj, 137, 4186
\bibitem{lap00} Lapasset, E., Claria, J. J., Mermilliod, J.-C., 2000, \aap, 361, 945	
\bibitem{lyn87} Lynga, G., Computer Based catalogue of Open Cluster Data, 5th. Ed. (Strassbourg: CDS, 1987)

\bibitem{math90} Mathis, J. 1990, ARA\&A, 28,37
\bibitem{mer81} Mermilliod, J. C., 1981,  \aaps, 44, 467
\bibitem{moi10} Moitinho, A., 2010, Star clusters: basic galactic building blocks, Proceedings IAU Symposium No.266, eds. R.de Grijs and J.R.D. Lepine 
\bibitem{pau14} Paunzen, E., Netopil, M., Maitzen, H. M., Pavlovski, K., Schnell, A., Zejda, M., 2014, \aap, 564, 42
\bibitem{pes61} Pesch, P., 1961, \apj, 134, 602
\bibitem{phe94} Phelps, R.L., Jane,s K.A., Montgomery, K.A., 1994, \aj, 107, 1079
\bibitem{roe10} Roeser, S., Demleitner, M., Schilbach, E., 2010, \aj, 139, 2440
\bibitem{sal04} Salaris, M., Weiss, A., Percival, S.M., 2004, \aap, 414, 163 
\bibitem{san69} Sandage, A., 1969, \apj, 158, 1115
\bibitem{sca92} Schaller, G., Shaerer, D., Meynet, G., Maeder, A., 1992, \aaps, 96, 269

\bibitem{sch98} Schlegel, D.J., Finkbeiner, D. P., and Davis, M., 1998, \apj, 500, 525
\bibitem{Sch87} Schneider, H., 1987,  \aaps, 71, 531	
\bibitem{sch82} Schmidt-Kaler, Th.\ 1982, in Landolt-Bornstein, Numerical Data and Functional Relationships in Science and Technology, New Series, Group VI, Vol.2b, eds.\ K.~Schaifers and H.~H. Voigt (Berlin:  Springer), p.~14 (SK82)
\bibitem{sch07} Schuster, W.~J., Michel, R., Dias, W., Tapia-Peralta, T., V\'azquez, R., Macfarland J., Chavarr\'{\i}a, C., Santos, C., and Moitinho, A. 2007, Galaxy Evolution Across the Hubble Time, eds.\ F.~Combes and J.~Palou$\breve{\rm s}$, Proceedings of the International Astronomical Union, IAU Symposium No.~235,(Cambridge, United Kingdom:  Cambridge University Press), p.~331
\bibitem{sha06} Sharma, S., Pandey, A. K., Ogura, K., Mito, H., Tarusawa, K., Sagar, R., 2006, AJ, 132, 1669

\bibitem{skr06} Skrutskie, M.F., Cutri, R., Stiening, R., Weinberg, M.D., Schneider, S.E., Carpenter, J.M., Beichman, C., Capps, R.,  2006, \aj, 131, 1163
\bibitem{stet87} Stetson P.~B., 1987, PASP, 99, 191

\bibitem{str95} Strai\c{z}ys, V. 1995, Multicolor Stellar Photometry, Astronomy and Astrophysics Series,
Vol.~15, ed.~A.~G.~Pacholczyk (Tucson, Arizona:  Pachart Pub.~House)

\bibitem{tap10} Tapia, M.~T., Schuster, W.~J., Michel, R., Chavarr\'ia-K, C., Dias, W.~S., V\'azquez, R., andMoitinho, A., 2010, \mnras, 401, 621 (T10)
\bibitem{twa97}Twarog, B.A., Ashman, K.M., Anthony-Twarog, B.J., 1997, \aj, 114, 2556
\bibitem{van01} van Dokkum P.~G., 2001, PASP, 113, 1420 
\bibitem{ven98} Ventura, P., Zeppieri, A., Mazzitelli, I., \& D’Antona, F., 1998,  \aap, 334, 953
\bibitem{yi01} Yi, S., Demarque, P., Kim, Y.-C., Lee, Y.-W., Ree, C.H., Lejeune, T., Barnes, S., 2001, \apj, 136, 417
\bibitem{zac13} Zacharias, N., Finch, C.T., Girard, T.M. \etal, 2013, \aj, 145, 44 

\end{theunbibliography}
\end{document}